**Title:**

Estimating common synaptic inputs to spinal motor neurons from motor unit spike trains using *openhdemg*

**Authors:**

Hélio V. Cabral[1,2,3,*], Giacomo Valli[4,*], Roberto Zanotti[4], Ioannis Delis[5], Francesco Negro[4]

**Co-first authors, as these authors contributed equally to this work.*

**Affiliations:**

[1]School of Physical Education and Sports, Universidade Federal do Rio de Janeiro, Rio de Janeiro, Brazil

[2]Biomedical Engineering Program (COPPE), Universidade Federal do Rio de Janeiro, Rio de Janeiro, Brazil

[3]Postgraduate Program in Rehabilitation Sciences, Universidade Federal do Rio de Janeiro, Rio de Janeiro, Brazil

[4]Department of Clinical and Experimental Sciences, Università di Brescia, Brescia, Italy

[5]School of Electrical and Computer Engineering, National Technical University of Athens, Athens, Greece

**Acknowledgements:**

Funded by the European Union. Views and opinions expressed are, however, those of the author(s) only and do not necessarily reflect those of the European Union or the European Research Council Executive Agency. Neither the European Union nor the granting authority can be held responsible for them. European Research Council Consolidator Grant INcEPTION n. 101045605 (F.N.).

**Corresponding author:**

Prof. Francesco Negro

Department of Clinical and Experimental Sciences

Università di Brescia

Viale Europa 11, Brescia, 25123, Italy

E-mail: francesco.negro@unibs.it

**Abstract**

Common synaptic input is considered a fundamental principle of motor neuron control and represents the dominant component of the neural drive transmitted from the motor neuron pool to muscle. Recent advances in High-Density surface Electromyography (HDsEMG) and motor unit (MU) decomposition algorithms have enabled the concurrent identification of increasingly large populations of MUs and substantially expanded the possibility of estimating common synaptic input from MU spike trains, making this approach widely used to investigate the neural control of movement in humans. However, multiple analytical approaches are currently available, each relying on different physiological assumptions, mathematical formulations, and parameter choices. Furthermore, the lack of practical guidelines and open-source implementations has limited the accessibility and reproducibility of these analyses. In this tutorial, we provide a practical, physiologically grounded guide to estimating common synaptic input from populations of MU spike trains using openhdemg, an open-source Python framework. We organize the available methods into three complementary categories: time-domain approaches applied to smoothed discharge rates, frequency-domain approaches based on coherence between cumulative spike trains, and a network-information approach based on nonlinear pairwise dependencies and graph theory. For each method, we describe its physiological interpretation, step-by-step estimation, and systematically examine how key parameter choices influence the resulting estimates, providing practical recommendations for their selection. Finally, we present a complete workflow from HDsEMG decomposition and MU cleaning to common synaptic input estimation, demonstrating that decomposition quality directly affects these estimates. This tutorial provides an accessible and reproducible framework for investigating common synaptic input from MU populations.

## 1. Introduction

Voluntary movement arises from the ability of the central nervous system to transform neural commands into coordinated muscle activation and, ultimately, muscle force production (Heckman and Enoka, 2012, Enoka and Farina, 2021). This transformation involves multiple interacting neural processes from descending, spinal, and sensory pathways that converge onto the alpha motor neuron pool, whose binary spike train output determines the neural command delivered to the active muscle (Ishizuka et al., 1979, Lemon, 2008, Duchateau and Enoka, 2011). In this sense, alpha motor neurons and the muscle fibers they innervate (i.e., motor units; MUs), represent the final integrative stage of the motor pathway through which synaptic inputs are transduced into muscle force (Liddell and Sherrington, 1925, Sherrington, 1925). The neural signal received by the muscle can then be approximated by the cumulative spike train, defined as the sum of the spike trains of the active MUs, and commonly referred to as the neural drive to the muscle (Farina and Negro, 2015).

The synaptic inputs integrated by alpha motor neurons include both common inputs, which are shared across motor neurons, and independent inputs, which differ across motor neurons. Because common inputs modulate the membrane potentials of multiple motor neurons simultaneously, they are enhanced in the neural drive to the muscle through a population averaging process, whereas independent inputs tend to be attenuated (Farina and Negro, 2015, Hug et al., 2023a). Accordingly, compelling evidence from both computational simulations and experimental studies has demonstrated that the motor neuron pool functions as a highly selective spatial filter, preserving synaptic inputs that are common to multiple motor neurons while reducing the influence of independent inputs (Negro et al., 2009, Negro and Farina, 2011b, Farina et al., 2014, Negro et al., 2016b). Together with the intrinsic low-pass filtering properties of muscle (Bawa and Stein, 1976, Baldissera et al., 1998, Cabral et al., 2024b), these findings indicate that the force output is primarily driven by the low-frequency components of synaptic input that are widely shared across the motor neuron pool. Collectively, these observations support the view that MU activity is primarily driven by low-dimensional neural control, with common synaptic input emerging as a central determinant of muscle force generation (Negro et al., 2009, Farina et al., 2014, Laine et al., 2015, Madarshahian et al., 2021, Del Vecchio et al., 2019, Del Vecchio et al., 2023, Levine et al., 2023, Rossato et al., 2024, Hug et al., 2023b, Dernoncourt et al., 2025).

An important consequence of common synaptic input being a major determinant of motor neuron output is that the discharge times of concurrently active MUs are not independent (Adrian and Bronk, 1928). Instead, MUs tend to exhibit coupled discharge behaviors, resulting in some degree of couplings among their outputs (Sears and Stagg, 1976, Datta and Stephens, 1990, Bremner et al., 1991, Nordstrom et al., 1992, De Luca et al., 1982, Mori, 1975). This phenomenon has motivated the use of correlation-based analyses between MU discharge times to infer the amount and structure of common input received by the motor neuron pool (Kirkwood, 1979). Since the 1970s, several measures have been proposed to quantify these relations in the time and frequency domains, including common drive index (De Luca et al., 1982, De Luca and Erim, 1994), short-term synchronization (Kirkwood and Sears, 1978, Nordstrom et al., 1992, McIsaac and Fuglevand, 2008, Farmer et al., 1993), and coherence analyses (Rosenberg et al., 1998, Farmer et al., 1993, Amjad et al., 1997, Rosenberg et al., 1989). Most early applications of these methods were based on pairwise analyses of MUs, which were justified by the relatively small number of MUs (~1-2) that could be identified concurrently during voluntary contractions using intramuscular wire or quadrifilar electrodes (Adrian and Bronk, 1928, Bigland and Lippold, 1954, De Luca and Forrest, 1972). Although pairwise analyses provided important physiological insights, they are intrinsically limited by the influence of motor neuron sampling properties on correlation estimates (de la Rocha et al., 2007, Lazar and Pnevmatikakis, 2008, Negro and Farina, 2012).

Building on pioneering work from the 1980s using multichannel electromyography (Masuda et al., 1983, Masuda et al., 1985, Reucher et al., 1987), advances in high-density electromyography and MU decomposition algorithms over the last decades (Merletti et al., 1999, Holobar and Zazula, 2007b, Ning et al., 2015, Muceli et al., 2015, Negro et al., 2016a, Chen and Zhou, 2016, Muceli et al., 2022) have enabled the identification of large populations of concurrently active MUs during voluntary contractions. This technological advance has substantially changed the way common synaptic input can be investigated (Farina et al., 2016). Rather than relying exclusively on pairs of MUs, the concurrent analysis of multiple MU spike trains can improve the estimation of shared inputs, because it increases the average sampling rate of the common input to alpha motor neurons (Negro and Farina, 2011a, Negro and Farina, 2011b). This principle underlies the use of cumulative spike trains and pooled estimates of coherence (Negro et al., 2016b, Del Vecchio et al., 2019, Maillet et al., 2022, Rossato et al., 2022, Cabral et al., 2024a, Cabral et al., 2024b, Cabral et al., 2025a), as well as dimensionality-reduction (Negro et al., 2009, Del Vecchio et al., 2023, Rossato et al., 2024, Dernoncourt et al., 2025, Cabral et al., 2025b) and, more recently, network-based approaches applied to MU discharge patterns (Hug et al., 2023b, Cabral et al., 2025b). Therefore, the availability of larger MU samples has not only improved the robustness of common input estimates but has also expanded the analytical framework from pairwise correlation to population-level estimates.

Several methods are currently available to estimate common synaptic input from MU spike trains, and each method provides a different physiological and mathematical representation of the shared input to the motor neuron pool. Time-domain approaches quantify shared fluctuations in smoothed discharge rates (De Luca et al., 1982, Negro et al., 2009, Del Vecchio et al., 2023, Rossato et al., 2024, Cabral et al., 2025b), frequency-domain approaches identify the spectral content of common input (Negro and Farina, 2012, Negro et al., 2016b), and more recent network-information theoretic approaches capture nonlinear dependencies among MU outputs (O'Reilly and Delis, 2024, O'Reilly and Delis, 2022, Cabral et al., 2025b). However, these methods are sensitive to several parameter selections, such as the duration of the smoothing window for time-domain approaches, the length and overlap of the analysis windows for frequency-domain approaches, and the number of MUs included in the analysis. These parameter choices can substantially influence the resulting estimates and, in some cases, lead to different physiological interpretations (Negro and Farina, 2012). Although tutorials and practical guidelines have been published for estimating MU discharge properties and analyzing central and peripheral MU features (Del Vecchio et al., 2020, Valli et al., 2024, Lecce et al., 2026), practical guidance on estimating common synaptic input from MU spike trains remains limited. Moreover, the absence of open-source implementations has often required researchers to rely on specialized knowledge in signal processing and advanced coding, which may limit the accessibility and reproducibility of this type of analysis across laboratories.

The purpose of this paper is to provide a practical and physiological guide for estimating common synaptic input from populations of MU spike trains using *openhdemg* (Valli et al., 2024) (https://www.giacomovalli.com/openhdemg/), an open-source framework written in Python 3 (Python Software Foundation, USA). We organize these approaches into three complementary categories according to the representation of MU activity used to estimate common synaptic input: (1) time-domain approaches applied to smoothed discharge rates, including the common drive index and principal component analysis (**Section 3**); (2) frequency-domain approaches based on coherence between cumulative spike trains, including pooled coherence and the proportion of common input index (**Section 4**); and (3) a network-information approach based on nonlinear pairwise dependencies and graph theory (**Section 5**). For each approach, we describe the physiological interpretation, the step-by-step estimation, and how key parameter choices affect the estimates, and provide practical recommendations for parameter selection. The supplementary material available at https://doi.org/10.6084/m9.figshare.32751243 complements these sections by providing implementation examples of *openhdemg*, including the main function calls and arguments required to reproduce each analysis. Finally, we present the complete workflow from HDsEMG decomposition to MU cleaning and common synaptic input estimation (**Section 6**), which can be performed within the *openhdemg* framework via an integrated, flexible Python-based interface. In particular, we discuss the effect of MU cleaning on the estimation of common synaptic input and demonstrate that the presence of a few poorly edited MUs within larger pools can alter these estimates.

## 2. Datasets used in this tutorial

To illustrate the implementation and parameter sensitivity of each approach, we used decomposed HDsEMG recordings from individual participants included in previously published studies. For the common drive index and pooled coherence analyses, we used HDsEMG signals recorded from the vastus lateralis of one participant from Cosentino et al. (2026). In that experiment, the participant performed an isometric knee extension learning task in which the target force was a randomly generated signal low-pass filtered at 1.5 Hz. The target oscillated around a force level corresponding to 10% of maximal voluntary contraction, and each trial lasted 30 s. The data used in the present tutorial were obtained from one of the last 15 trials, after the participant had learned the task. For the principal component analysis of smoothed discharge rates and the network-information framework, we used data from another participant included in Cabral et al. (2025b). In this dataset, HDsEMG signals were recorded from the tibialis anterior while the participant performed isometric dorsiflexion contractions at 10% of maximal voluntary contraction. The task followed a similar structure to that described for the vastus lateralis dataset, with the participant tracking a randomly generated force target. Finally, for the proportion of common input index and for the analysis of how MU cleaning affects estimates of common synaptic input, we used data from one participant included in Cabral et al. (2024b). In that experiment, the participant performed an isometric dorsiflexion task at 10% of maximal voluntary contraction for 40 s, with the ankle positioned at 130°. Together, these datasets were selected to provide representative examples of decomposed MU spike trains from different muscles, contraction tasks, and analysis scenarios within the *openhdemg* framework.

The final MU yield was 18 MUs for the vastus lateralis dataset used for the common drive index and pooled coherence analyses, 21 MUs for the tibialis anterior dataset used for the principal component analysis and network-information framework, and 31 MUs for the tibialis anterior dataset used for the proportion of common input index and for the analyses presented in **Section 6**. The only required input for all functions used to estimate common synaptic input is the *emgfile*, which is the Python object containing the information from the decomposed HDsEMG file and represents the basic data structure of the *openhdemg* framework. This object includes, among other variables, the binary MU spike trains, the sampling frequency, and the number of identified MUs. Therefore, starting from the binary spike trains contained in the *emgfile*, the functions automatically perform the main processing steps described in the following sections. Further details about the *openhdemg* data structure are available at https://www.giacomovalli.com/openhdemg/tutorials/emgfile_structure/.

## 3. Time-domain approaches

### 3.1. Common drive index

#### *3.1.1. Physiological interpretation*

The common drive index quantifies the extent to which the discharge rates of MUs share low-frequency fluctuations over time. These shared fluctuations are assumed to reflect low-frequency common synaptic input to the motor neuron pool, commonly termed common drive (De Luca et al., 1982, De Luca and Erim, 1994). The method is based on pairwise cross-correlation analysis of MU discharge rates, with the common drive index typically defined as the average cross-correlation across all MU pairs. This approach was initially applied to MUs recorded from the deltoid and first dorsal interosseous muscles (De Luca et al., 1982) and continues to be widely used in the MU literature (Hug et al., 2023b, Levine et al., 2023, Maillet et al., 2022). However, although the common drive index provides a useful estimate of shared low-frequency modulation among MUs, it is important to recognize that the strength of common synaptic input received by a pair of motor neurons is not necessarily proportional to the degree of correlation between their output spike trains (de la Rocha et al., 2007, Negro and Farina, 2012, Rodriguez-Falces et al., 2017). Therefore, even though the common drive index can be also interpreted as a population-level estimate because it averages cross-correlations across all possible MU pairs, recent studies have adopted a more conservative interpretation by focusing on the proportion of MU pairs whose cross-correlation values exceed a statistical confidence threshold (Hug et al., 2023b, Levine et al., 2023, Maillet et al., 2022).

*3.1.2. Step-by-step estimation*

The first step in estimating the common drive index is to obtain the smoothed discharge rate of each MU from its binary spike train (**Figure 1A**). Let $x_i(t)$ denote the binary spike train of the $i^{th}$ MU, where $x_i(t) = 1$ when the MU discharges and $x_i(t) = 0$ otherwise. The smoothed discharge rates $y_i(t)$ is classically obtained by convolving $x_i(t)$ with a smoothing window $w(t)$:

$$y_i(t) = x_i(t) * w(t) = \int_{-\infty}^{+\infty} x_i(\tau) w(t - \tau) d\tau \quad (1)$$

The smoothing kernel typically used for this purpose is a 0.4-s Hanning window (De Luca et al., 1982), which corresponds to low-pass filtering the MU discharge rate at ~1.8 Hz (Negro et al., 2009). This operation emphasizes slow fluctuations in MU discharge rates and attenuates higher-frequency fluctuations, thereby isolating the low-frequency oscillations assumed to reflect common drive. Because convolution and subsequent filtering can introduce edge artifacts, a portion of the signal at the beginning and end of the smoothed discharge rates can be removed before further analysis, for example 1 s at each edge. Subsequently, a zero-phase high-pass filter with a cut-off frequency of 0.75 Hz is applied to remove offsets and slow trends from the smoothed discharge rates in the smoothed discharge rates (De Luca et al., 1982), resulting in the detrended smoothed discharge rates shown in **Figure 1B**.

The next step is to quantify the similarity between the smoothed discharge rates of each pair of MUs. For the $i^{th}$ and $j^{th}$ MUs, the normalized cross-correlation $\rho_{ij}$ is calculated as a function of a time lag $\tau$:

$$\rho_{ij}(\tau) = \frac{\gamma_{ij}(\tau)}{\sqrt{\gamma_{ii}(0)\gamma_{jj}(0)}} \tag{2}$$

where $\gamma_{ij}(\tau)$ is the cross-covariance between $y_i(t)$ and $y_j(t)$ at a lag $\tau$, $\gamma_{ii}(0)$ and $\gamma_{jj}(0)$ are the zero-lag auto-variances of $y_i(t)$ and $y_j(t)$, respectively. This normalization constrains the cross-correlation values between -1 and 1, allowing the strength of the association between MU discharge rate fluctuations to be compared across pairs. The cross-correlation functions obtained for all MU pairs are shown as gray lines in **Figure 1C**. From each cross-correlation function, the peak correlation value within a physiological lag range (typically ± 0.1 s; De Luca et al. (1982)) is extracted and used as the pairwise estimate of common drive. These pairwise peak correlation values can then be arranged into a correlation matrix of $N\ x\ N$ MUs (**Figure 1D**). The normalized cross-correlation can be calculated using the entire smoothed discharge rate signal or after segmenting the signal into shorter time windows, for example 5-s windows (Negro et al., 2009). When windowed analysis is used, the peak cross-correlation is first extracted for each MU pair within each time window and then averaged across windows to obtain a single pairwise correlation value. Finally, the common drive index is computed as the average of the pairwise peak correlation values across all possible MU pairs. Therefore, the common drive index provides a summary measure of the extent to which the population of MUs exhibits low-frequency fluctuations in discharge rates over time.

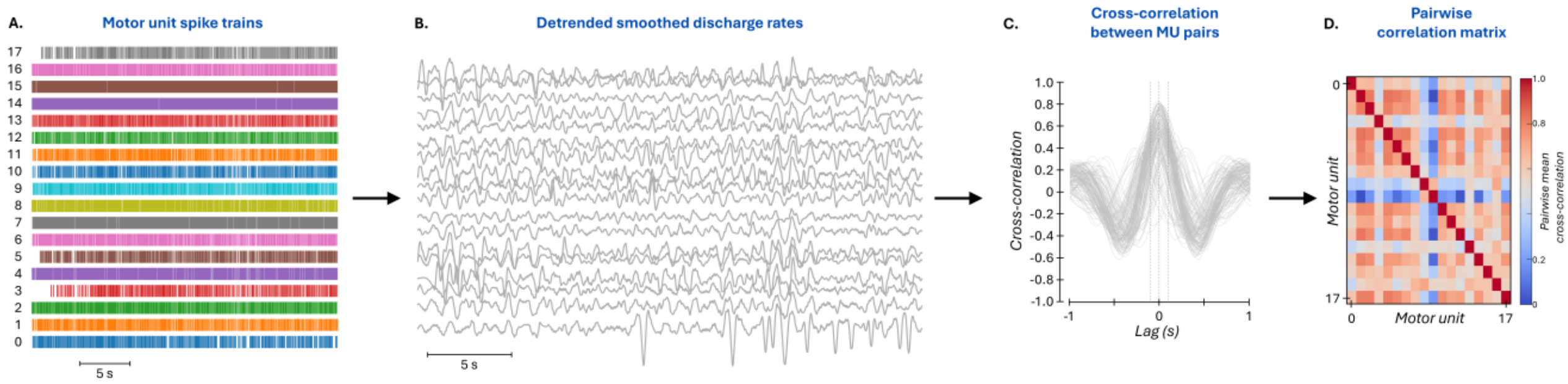

***Figure 1: Step-by-step estimation of the common drive index from motor unit spike trains.*** *(A) Motor unit spike trains obtained from decomposed high-density surface electromyography (HDsEMG) signals. (B) Smoothed discharge rates obtained by convolving the motor unit spike trains with a 0.4-s Hanning window, followed by detrending using a zero-phase high-pass filter with a cutoff frequency of 0.75 Hz. (C) Normalized cross-correlation functions calculated for all possible motor unit pair combinations using the detrended smoothed discharge rates shown in panel B. Peak cross-correlation values were extracted within a lag window of ± 0.1 s (vertical dashed lines). (D) Pairwise correlation matrix constructed from the peak cross-correlation values of all motor unit pairs. In this example, the matrix has dimensions of 18 × 18 motor units. The common drive index is calculated as the average of the peak cross-correlation values across all motor unit pairs. The color scale represents the strength of the correlation between motor unit pairs.*

### *3.1.3. Statistical significance*

A statistical confidence level can be calculated to determine whether the peak cross-correlation values observed between MU pairs exceed the level expected by chance. In recent studies, this has typically been achieved using surrogate spike trains generated by randomly shuffling the interspike intervals of the original MU discharge times (Hug et al., 2023b, Levine et al., 2023, Maillet et al., 2022). This procedure preserves the discharge statistics of each MU (e.g., the number of discharges and the interspike interval distribution), while disrupting the temporal relationship between MUs and therefore reducing the correlations expected from common synaptic input. After generating the surrogate spike trains, the same processing pipeline applied to the original data is repeated. Specifically, surrogate spike trains are converted into smoothed discharge rates, edge portions affected by convolution and filtering are removed, the signals are detrended, and pairwise normalized cross-correlations are calculated using the same analysis windows and lag range described in the previous section. When two surrogate versions are generated for each MU, each real MU pair can be compared with four possible combinations of surrogate spike trains. These surrogate repetitions are then used to build a null distribution of peak cross-correlation values expected in the absence of temporal coupling between MUs. The confidence level is then defined as a high percentile of this surrogate distribution, commonly the 95th or 99th percentile. Pairwise peak cross-correlation values from the original MU spike trains that exceed this confidence level are considered statistically significant. This approach for calculating the confidence level of the pairwise cross-correlation values is the one implemented in *openhdemg*.

### 3.1.4. Parameter sensitivity and recommendations

The estimation of the common drive index is implemented in *openhdemg* through the function *common_drive_index*, with the main structure described in Supplementary Material. We systematically tested how three main parameters affected the estimates (**Figure 2**): the duration of the time windows used to segment the smoothed discharge rates for cross-correlation analysis, the duration of the Hanning window used to smooth the binary spike trains, and the overlap between consecutive windows used for cross-correlation analysis.

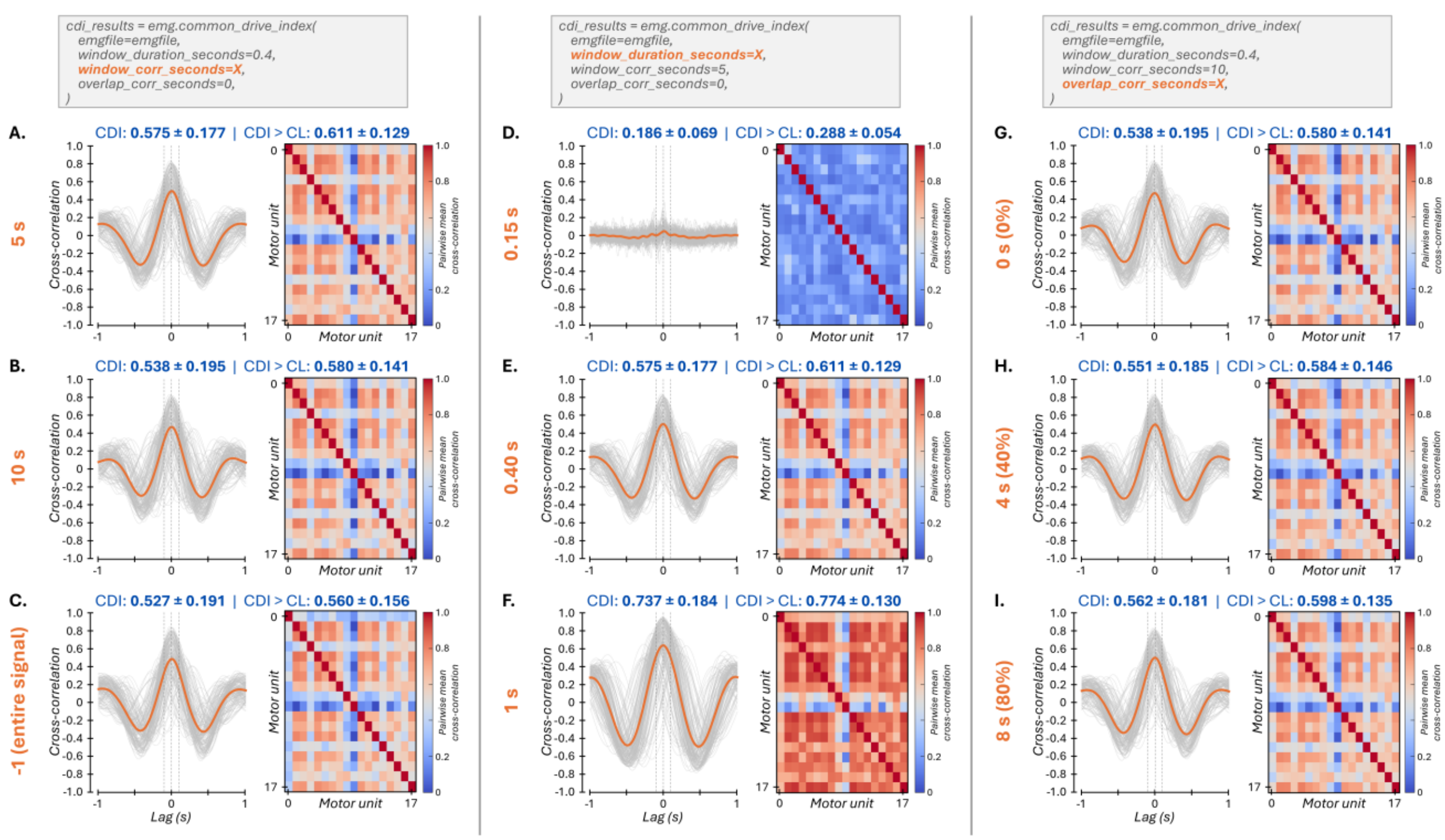

***Figure 2: Parameter sensitivity of the common drive index.*** *The common drive index was estimated from detrended smoothed motor unit discharge rates using pairwise normalized cross-correlation analysis. Panels A–C show the effect of changing the duration of the time windows used to segment the smoothed discharge rates for cross-correlation analysis: 5-s windows (A), 10-s windows (B), and the entire signal (C). Increasing the duration of the cross-correlation window reduced the estimated common drive index, indicating that shorter windows are more sensitive to local periods of shared discharge-rate modulation. Panels D–F show the effect of changing*

*the duration of the Hanning window used to smooth the binary spike trains: 0.15 s (D), 0.4 s (E), and 1 s (F). Longer smoothing windows increased the common drive index by attenuating high-frequency discharge variability and emphasizing slower fluctuations shared across motor units, although this progressively restricts the analysis to lower-frequency components of motor unit discharge rates. Panels G–I show the effect of changing the overlap between consecutive cross-correlation windows while using 10-s windows: 0 s overlap (G), 4 s overlap (H), and 8 s overlap (I). Increasing the overlap produced modestly higher and smoother estimates of common drive, but adjacent overlapping windows are not statistically independent. In each panel, the correlation matrix represents the pairwise peak cross-correlation values between motor units, with higher values (red) indicating stronger shared low-frequency modulation.*

First, we tested the effect of the duration of the cross-correlation window by segmenting the smoothed MU discharge rates into 5-s windows (**Figure 2A**), 10-s windows (**Figure 2B**), or by using the entire signal for the cross-correlation analysis (**Figure 2C**). Increasing the duration of the cross-correlation window progressively reduced the estimated common drive index, as reflected by the lower pairwise correlation values in the MU correlation matrix (note that there are more blue colors from **Figure 2A** to **Figure 2C**). Note also that, in the example shown, the common drive index decreased from 0.58 ± 0.18 for the 5-s window segmentation to 0.53 ± 0.19 for the entire signal. Given the inherent non-stationarity of MU discharge rates, shorter analysis windows improve the temporal resolution needed to capture time-varying adjustments in common fluctuations. Furthermore, the slight reduction in standard deviation observed with the 5-s windows suggests that segmentation provides a more stable, localized profile of MU behavior.

Second, we tested the effect of the duration of the Hanning window used to smooth the binary spike trains. We systematically compared smoothing windows of 0.15 s (**Figure 1D**), 0.4 s (**Figure 1E**), and 1 s (**Figure 1F**), which are approximately equivalent to low-pass filtering the MU discharge rates with -3 dB cutoff frequencies of ~4.8 Hz, ~1.8 Hz, and ~0.72 Hz, respectively. Increasing the duration of the smoothing window markedly increased the common drive index. In the example shown, the common drive index increased from 0.19 ± 0.07 for the 0.15-s window to 0.74 ± 0.18 for the 1-s window. Similarly, when only significant MU pairs were considered, the thresholded common drive index increased from 0.29 ± 0.05 to 0.77 ± 0.13. This behavior is expected because longer smoothing windows attenuate more high-frequency discharge rate fluctuations and emphasize slower fluctuations that are more likely to be shared across MUs, as they are relevant for muscle force production (Cabral et al., 2025b, Negro et al., 2009, Thompson et al., 2018). However, this increase should not be interpreted simply as stronger common synaptic input. Rather, it reflects the frequency content emphasized by the smoothing procedure. Shorter smoothing windows preserve better temporal resolution but allow more high-frequency variability to remain in the discharge rate estimates, whereas longer windows provide stronger noise attenuation at the expense of excessively smoothing the signals. Thus, although a 1-s Hanning window may produce higher common drive index values, the resulting estimate primarily reflects very low-frequency fluctuations in MU discharge rates.

Third, we tested the effect of overlapping between consecutive cross-correlation windows. For this analysis, the duration of the cross-correlation window was fixed at 10 s to better illustrate the effect of overlapping windows. We compared 0 s overlap (corresponding to non-overlapping windows; **Figure 2G**), 4 s overlap (corresponding to 40% overlap; **Figure 2H**), and 8 s overlap (corresponding to 80% overlap: **Figure 2I**). Increasing the overlap between windows produced a modest increase in the common drive index from 0.54 ± 0.20 for non-overlapping to 0.56 ± 0.18 for 80% overlap. This occurred because overlapping windows provide a smoother sampling of the time-varying cross-correlation structure and reduce abrupt changes between consecutive segments (i.e., more stable estimates). However, overlapping windows are not statistically independent, because adjacent windows share part of the same data. Therefore, although overlap can stabilize the estimate, it should not be interpreted as adding independent information to the analysis.

Based on these observations, a 5-s cross-correlation window is a practical default for estimating the common drive index during steady contractions, because it provides sensitivity to local shared fluctuations while maintaining sufficient data within each window for cross-correlation analysis. For smoothing the binary spike trains, a 0.4-s Hanning window represents a reasonable compromise between preserving temporal resolution and isolating the low-frequency fluctuations classically associated with common drive. Finally, using non-overlapping windows as the default approach allows to avoid overweighting redundant portions of the signal. For longer contractions, however, it may be useful to test the effect of window overlapping. In all cases, parameter choices should be reported explicitly, because changes in smoothing duration, cross-correlation window duration, and window overlapping can influence the magnitude and physiological interpretation of the common drive index.

### 3.2. Principal component analysis of smoothed discharge rates

#### *3.2.1. Physiological interpretation*

Principal component analysis (Hotelling, 1936, Hotelling, 1933, Pearson, 1901) applied to smoothed discharge rates provides a data-driven approach to identify the dominant low-dimensional patterns shared across a population of MUs (Negro et al., 2009, Cabral et al., 2025b). Dimensionality reduction techniques, such as principal component analysis and factor analysis, have been widely applied to neuronal ensemble activity to extract dominant patterns of covariation and identify a reduced set of components that capture relevant features of neural control (Cunningham and Yu, 2014). When applied to smoothed MU discharge rates, these methods allow the investigation of how much of the population activity can be explained by a small number of common components. Unlike the common drive index, which provides a scalar summary of the average pairwise association among MUs, principal component analysis provides time-varying signals representing the dominant modes of covariation in the MU population. In this context, the first principal component represents the linear combination of smoothed discharge rates that explains the largest amount of shared variance across MUs (Negro et al., 2009). Therefore, when MU activity is strongly governed by common synaptic input, a large proportion of the total variance is expected to be captured by the first principal component or by a small number of retained components. The application of dimensionality reduction to MU spike trains was first introduced to investigate the low-dimensional organization of MU discharge rates, revealing that the first principal component of MU discharge rates can explain oscillations in isometric force more effectively than the discharge rates of individual MUs (Negro et al., 2009). Since then, several studies have used principal component analysis or related approaches to estimate low-dimensional components that explain the variability of smoothed discharge rates in populations of active MUs (Del Vecchio et al., 2023, Rossato et al., 2024, Nuccio et al., 2024, Dernoncourt et al., 2025, Cabral et al., 2025b). This finding supports the physiological interpretation that the dominant component extracted from the population of MU discharge rates may reflect the common synaptic input transmitted to the motor neuron pool and expressed in force output.

*3.2.2. Step-by-step estimation*

The initial steps for estimating the low-dimensional components underlying MU discharge rates are similar to those used for the common drive index. The smoothed discharge rate $y_i(t)$ of the $i^{th}$ MU is obtained from its binary spike train $x_i(t)$ (**Figure 3A**) as described in equation (1). As for the common drive index, a 0.4-s Hanning window is typically used to obtain the smoothed discharge rates (Negro et al., 2009). Edge portions of the signal affected by convolution and filtering can then be removed, for example 1 s at the beginning and end of the signal. Subsequently, offsets and slow trends are removed from the smoothed discharge rates using a zero-phase high-pass filter with a cut-off frequency of 0.75 Hz (**Figure 3B**).

After preprocessing, the smoothed discharge rates from all MUs are organized into a data matrix $Z$, with dimension $T \ x \ N$, where $T$ is the number of time samples, and $N$ is the number of MUs. Before applying principal component analysis, each MU discharge rate in this matrix is typically standardized to have zero mean and unit variance (Joliffe and Morgan, 1992). This step ensures that all MUs contribute to the analysis on the same scale and prevents MUs with larger discharge-rate variance from dominating the estimation of the principal components. After standardization, the covariance matrix is equivalent to the correlation matrix of the smoothed discharge rates. The next step is to compute the symmetric covariance matrix of the standardized smoothed discharge rates:

$$\Sigma = \begin{pmatrix} \gamma_{11} & \cdots & \gamma_{1N} \\ \vdots & \ddots & \vdots \\ \gamma_{N1} & \cdots & \gamma_{NN} \end{pmatrix} \tag{3}$$

where $\gamma_{ii}$ represents the variance of the standardized smoothed discharge rate of the $i^{th}$ MU, and $\gamma_{ij}$ represents the covariance between the standardized smoothed discharge rates of the $i^{th}$ and $j^{th}$ MUs. Because the signals are standardized, the diagonal elements of Σ are equal to one, and the off-diagonal elements represent the pairwise correlations between MU discharge rate fluctuations. The covariance matrix is then subject to eigenvalue decomposition:

$$\Sigma = V\Lambda V^T \tag{4}$$

where $V$ is an $N\ x\ N$ matrix containing the eigenvectors as columns $(v_1, v_2, \dots, v_N)$, and Λ is an $N\ x\ N$ diagonal matrix containing the corresponding eigenvalues $(\lambda_1, \lambda_2, \dots, \lambda_N)$, sorted in descending order. Each eigenvector contains $N$ values of the weights (or loadings) of each MU. Each eigenvector defines a weighted combination of MUs, whereas each eigenvalue quantifies the amount of variance in the population of smoothed discharge rates explained by the corresponding component. Because the standardized data has a total variance equal to the number of MUs $N$, the percentage of variance explained by the $k^{\text{th}}$ component is given by:

$$\%\ Variance\ by\ PC_k = \frac{\lambda_k}{N} \times 100 \tag{5}$$

The next step is to decide the number $k$ of components to retain (**Figure 3C**). This is a fundamental decision in principal component analysis and is covered in detail in the next subsection. After selecting $k$, only the first $k$ columns of $V$ are retained, resulting in the truncated eigenvector matrix $V_k$, with dimensions $N\ x\ K$. Finally, the standardized and detrended smoothed discharge-rate matrix is multiplied by $V_k$ to obtain the time-varying low-dimensional MU components:

$$Y = \mathrm{Z}V_K \tag{6}$$

where $Y$ is the $TxK$ matrix containing the retained principal component time series (**Figure 3D**). These components represent the dominant low-dimensional patterns shared across the population of MU discharge rates.

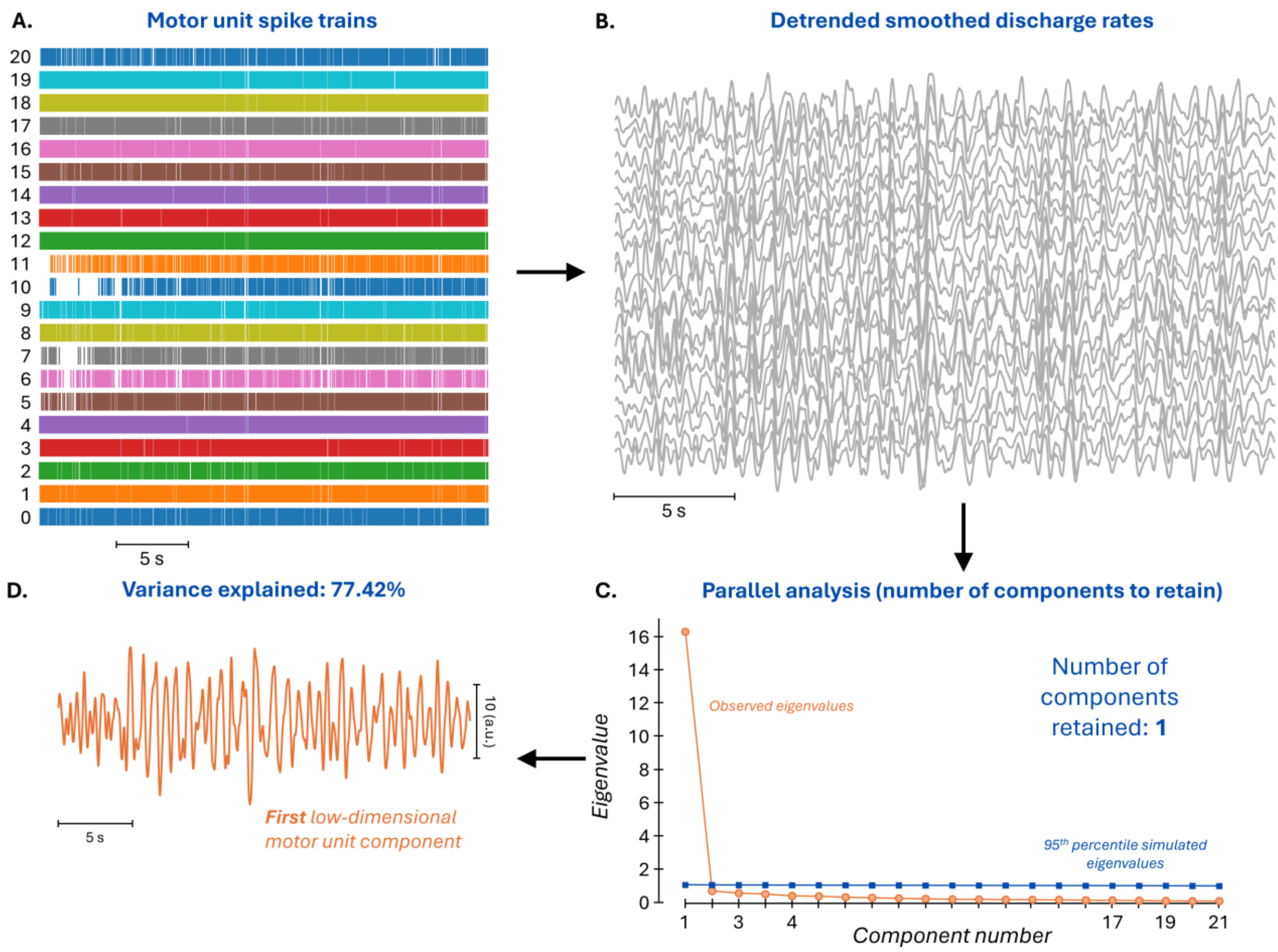

***Figure 3: Step-by-step estimation of principal component analysis (PCA) from motor unit spike trains.*** *(A) Motor unit spike trains obtained from decomposed high-density surface electromyography (HDsEMG) signals. (B) Smoothed discharge rates obtained by convolving the motor unit spike trains with a 0.4-s Hanning window, followed by detrending using a zero-phase high-pass filter with a cutoff frequency of 0.75 Hz. (C) Parallel analysis used to determine the number of low-dimensional components to retain. Observed eigenvalues (orange) are compared with the 95th percentile of eigenvalues obtained from randomized datasets (blue). In this example, only the first eigenvalue exceeded the parallel analysis threshold, indicating that a single component should be retained. (D) First low-dimensional motor unit component extracted from the detrended smoothed discharge rates. This component represents the dominant shared fluctuation across the motor unit population and explains the largest proportion of variance in the motor unit discharge rates.*

### *3.2.3. Number of components to retain*

An important step when applying principal component analysis (or other data dimensionality reduction approaches) to MU discharge rates is determining the number of low-dimensional components to retain. Several approaches have been proposed for this purpose, many of them based on the eigenvalues obtained from the decomposition of the covariance or correlation matrix (Guttman, 1954, Horn, 1965, Kaiser, 1960, Cattell, 1966, Velicer, 1976). One commonly used criterion is the $\lambda > 1$ (eigenvalue-greater-than-one) rule, also known as the Kaiser criterion (Kaiser, 1960). When principal component analysis is applied to standardized MU discharge rate data, the total variance of the population is equal to the number of MUs ($N$). Therefore, components with eigenvalues greater than one explain more variance than a single original MU discharge rate and are retained. Although this criterion is simple and intuitive, it may overestimate or underestimate the number of relevant components depending on the structure of the data and the number of MUs included in the analysis. Another commonly used approach is based on a cumulative variance threshold. In this case, components are retained until a predefined percentage of the total variance is explained (e.g., 70%, 80%, or 90%). This method is useful because it directly quantifies how much of the variability in the population of smoothed discharge rates is preserved by the retained low-dimensional representation. However, the choice of the variance threshold is arbitrary and may strongly influence the number of components retained. An alternative data-driven approach is parallel analysis (Horn, 1965, Zwick and Velicer, 1986, Velicer et al., 2000, Hayton et al., 2004), which has been considered one of the most accurate methods for identifying the number of components to retain and has been shown to outperform several traditional approaches. A detailed tutorial on the use of parallel analysis is provided by Hayton et al. (2004). Briefly, in the context of MU discharge rates, parallel analysis compares the eigenvalues from the real data with those expected from random data of similar dimension. To generate the random data, the time samples of each standardized and detrended smoothed discharge-rate signal are randomly shuffled. This procedure preserves the distribution of each MU signal but destroys the temporal covariation among MUs. Principal component analysis is then applied to this randomized data matrix, and the eigenvalues obtained for each component are stored. This procedure is repeated for several iterations (e.g., 1,000 times), resulting in a distribution of random eigenvalues for each component order. The number of low-dimensional MU components to retain is then determined by comparing the eigenvalues extracted from the real data with the eigenvalues expected from the randomized data. Specifically, a component is retained if its eigenvalue exceeds the upper bound of the random eigenvalue distribution, typically defined as the $95^{th}$ percentile. All three approaches for selecting the number of low-dimensional components are implemented in *openhdemg*, but parallel analysis is the default as it provides a data-driven estimate of the dimensionality of MU discharge-rate patterns while reducing the dependence on arbitrary variance or eigenvalue thresholds.

### 3.2.4. Parameter sensitivity and recommendations

The estimation of the low-dimensional MU components is implemented in *openhdemg* through the function *smoothed_dr_pca*, with the main structure described in Supplementary Material. For this approach, we systematically tested how three main parameters affected the estimates (**Figure 4**): the number of MUs included in the analysis, the duration of the Hanning window used to smooth the binary spike trains, and the method used to select the number of low-dimensional components to retain.

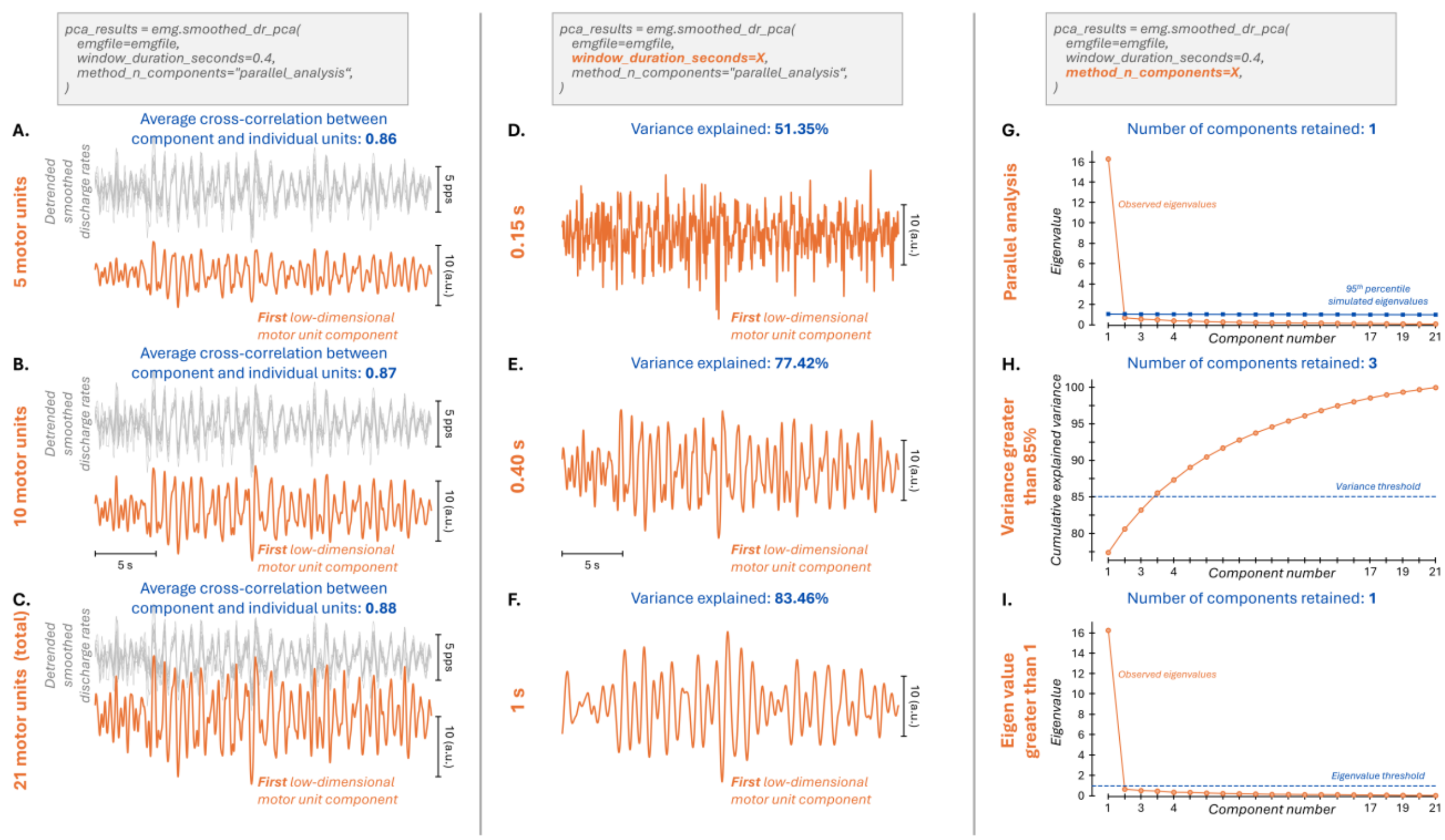

***Figure 4: Parameter sensitivity of principal component analysis applied to smoothed motor unit discharge rates.*** *Principal component analysis was applied to standardized and detrended smoothed discharge rates to extract the dominant low-dimensional components shared across the motor unit population. Panels A–C show the effect of changing the number of motor units included in the analysis: 5 motor units (A), 10 motor units (B), and all 21 motor units available in the dataset (C). Gray lines represent the smoothed discharge rates of all individual motor units, whereas the orange line represents the first low-dimensional component. The first component already captured the dominant shared fluctuations when 5 motor units were used, with only subtle increases in the average cross-correlation between the component and individual motor unit discharge rates when 10 or 21 motor units were included. Panels D–F show the effect of changing the duration of the Hanning window used to smooth the binary spike trains: 0.15 s (D), 0.4 s (E), and 1 s (F). Longer smoothing windows reduced high-frequency discharge variability and increased the percentage of variance explained by the first component, but progressively restricted the analysis to slower fluctuations. Panels G–I show the effect of changing the*

*criterion used to select the number of components to retain: parallel analysis (G), cumulative variance threshold greater than 85% (H), and eigenvalue-greater-than-one criterion (I). Parallel analysis and the eigenvalue-greater-than-one criterion retained one component in this dataset, whereas the cumulative variance criterion retained three components.*

First, we tested the effect of the number of MUs included in the analysis. Although this is not a specific input parameter of the function implemented in *openhdemg*, it is an important practical factor because the estimation of population-level components depends on the number of MUs available from the decomposition. To explore this effect, we randomly selected 5 MUs (**Figure 4A**), 10 MUs (**Figure 4B**), and used the full population of 21 MUs available in the dataset (**Figure 4C**) to estimate the low-dimensional MU components. When only 5 MUs were used, the first low-dimensional component already followed well the smoothed discharge rates of the MUs, showing an average cross-correlation of 0.86. When 10 MUs or all 21 MUs were used, the first low-dimensional component showed a smoother and more representative oscillatory profile, but the increases in average cross-correlation were subtle, reaching 0.87 and 0.88, respectively. These results indicate that a relatively small number of MUs may already be sufficient to capture the dominant shared fluctuations in the MU population. However, this might be muscle-dependent and increasing the number of MUs provides a more representative estimate of population activity.

Second, we tested the effect of the duration of the Hanning window used to smooth the binary spike trains. As for the common drive index, we compared smoothing windows of 0.15 s (**Figure 4D**), 0.4 s (**Figure 4E**), and 1 s (**Figure 4F**). With the 0.15-s window, the first low-dimensional component contained more high-frequency fluctuations and appeared visually noisier, explaining 51.35% of the variance in the smoothed discharge rates. Increasing the smoothing duration produced smoother components and increased the percentage of variance explained by the first component to 77.42% with the 0.4-s window and 83.46% with the 1-s window. This increase is expected because longer smoothing windows attenuate more high-frequency variability and emphasize slower fluctuations that are more likely to be shared across MUs. However, as discussed for the common drive index, a higher percentage of explained variance should not automatically be interpreted as a better or stronger estimate of common synaptic input. Longer smoothing windows progressively restrict the analysis to lower-frequency fluctuations. Therefore, although a 1-s Hanning window may produce a larger variance explained by the first component, the resulting component mainly reflects very low-frequency modulation of MU discharge rates, at the expense of temporal resolution.

Third, we tested the effect of the method used to select the number of components to retain. We compared parallel analysis (**Figure 4G**), a cumulative variance threshold criterion (**Figure 4H**), and the eigenvalue-greater-than-one criterion (**Figure 4I**). In this example, both parallel analysis and the eigenvalue-greater-than-one criterion retained a single component. Although both methods led to the same result in this dataset, parallel analysis provides a more data-driven criterion because the eigenvalues from the real data are compared with those expected from randomized data from the random shuffling of the original data. In contrast, the eigenvalue-greater-than-one criterion relies on a fixed threshold and may over- or underestimate the number of meaningful components. When the cumulative variance threshold was set to 85%, three components were retained. Although this approach ensures that a predefined amount of total variance is represented, it may retain additional components that explain residual variance but are not necessarily physiologically meaningful in terms of common synaptic input (Cabral et al., 2025b). Therefore, the choice of component-retention method can substantially affect the dimensionality inferred from the MU population.

Based on these observations, using as many well-identified and sufficiently cleaned MUs as available may provide a more representative estimate of population activity. However, an extremely large number of MUs may not always be necessary to extract the dominant low-dimensional component, particularly when the first component is strongly shared across the population. For smoothing, a 0.4-s Hanning window is a practical default, consistent with the common drive index and with previous applications of dimensionality reduction to MU discharge rates. This duration provides a reasonable compromise between attenuating high-frequency discharge variability and preserving the temporal structure of low-frequency common fluctuations. Finally, parallel analysis should be used as the default method for selecting the number of components to retain, as it provides a data-driven criterion and reduces dependence on arbitrary eigenvalue or cumulative variance thresholds. In all cases, the number of MUs, the smoothing window duration, the component-retention method, and the percentage of variance explained by the retained components should be explicitly reported.

## 4. Frequency-domain approaches

### 4.1. Pooled coherence

#### *4.1.1. Physiological interpretation*

Pooled coherence quantifies the frequency-specific coupling between populations of MU spike trains. Unlike time-domain approaches, which estimate shared fluctuations in smoothed discharge rates, coherence analysis identifies the spectral content of the common synaptic input to the motor neuron pool. Therefore, significant coherence between MU spike trains at a given frequency is interpreted as evidence that motor neurons receive common oscillatory synaptic input at that frequency (Castronovo et al., 2015, Negro and Farina, 2012, Negro et al., 2016b, Del Vecchio et al., 2019). In this context, coherence analysis has shown that MU spike trains may exhibit significant coupling across a broad range of frequencies up to approximately 80 Hz during isometric contractions (Castronovo et al., 2015, McManus et al., 2019, Muceli et al., 2022). The interpretation of coherence is frequency dependent. Coherence in the delta band, typically defined between 1 and 5 Hz, has been associated with the effective low-frequency control signal delivered to the motor neuron pool, as this frequency range overlaps with the bandwidth of force fluctuations (Farina et al., 2014). Coherence in the alpha band, typically defined between 5 and 15 Hz, has often been associated with physiological tremor and rhythmic force fluctuations observed during sustained contractions (Lippold, 1971, McAuley and Marsden, 2000). These oscillations may partly reflect afferent inputs from the active muscles projecting to the motor neuron pool (Hagbarth and Young, 1979, Christakos et al., 2006, Laine et al., 2016, Cabral et al., 2024b). Coherence in the beta band, typically defined between 15 and 35 Hz, is commonly interpreted as reflecting corticospinal contributions to motor neuron activity, because coupling between cortical and muscle activity has been observed within this frequency range (Conway et al., 1995, Baker, 2007, Baker et al., 1997, Negro and Farina, 2011b). Finally, gamma-band coherence, commonly defined as frequencies above 35 Hz, may reflect higher-frequency components of common synaptic input, including fast descending or spinal inputs, although its estimation from MU spike trains is more challenging and requires careful interpretation.

An important physiological and methodological consideration is that the relation between synaptic input to motor neurons and their output spike trains is nonlinear (Negro and Farina, 2011a, Negro and Farina, 2011b). Motor neurons transform continuous synaptic inputs into discrete discharge times, and this sampling-like process becomes particularly limited at higher frequencies because individual MUs discharge at relatively low rates, often around 10-20 Hz during low-to-moderate force contractions (Inglis et al., 2025). As a result, estimating coherence at higher frequencies, particularly in the beta and gamma bands, generally requires larger samples of concurrently active MUs to improve the effective sampling of common input and reduce estimation variability (Lazar and Pnevmatikakis, 2008, Negro and Farina, 2011a, Negro and Farina, 2011b). This is one of the main reasons why pooled approaches, based on cumulative spike trains from groups of MUs, are preferred over pairwise coherence when the goal is to characterize the spectral structure of common synaptic input to the motor neuron pool (Farina and Negro, 2015). Therefore, pooled coherence provides a population-level estimate of oscillatory common synaptic input.

*4.1.2. Step-by-step estimation*

In contrast to the time-domain approaches described above, pooled coherence is not calculated from smoothed discharge rates. Instead, this method is applied directly to the binary spike trains of MUs (**Figure 5A**). The first step is to split the population of MU spike trains into two non-overlapping groups of equal size, each commonly containing half of the total number of available MUs (Cabral et al., 2025a, Cabral et al., 2024a, Cabral et al., 2024b) (**Figure 5B**). To reduce the dependency of the coherence estimates on a specific allocation of motor units, the assignment of motor units to the two groups is performed randomly. For each group, the binary spike trains are summed to obtain two cumulative spike trains, $x(t)$ and $y(t)$ (**Figure 5B**). The use of cumulative spike trains increases the effective sampling of the common synaptic input received by the motor neuron pool, thereby improving the robustness of common input estimates compared with analyses based on individual MU pairs (Negro and Farina, 2011a, Negro and Farina, 2011b). The next step is to estimate the auto-spectrum of $x(t)$, the auto-spectrum of $y(t)$, and the cross-spectrum between $x(t)$ and $y(t)$. These spectra can be defined as the Fourier transforms of the corresponding auto- and cross-correlation functions:

$$P_{xx}(f) = \int_{-\infty}^{+\infty} \rho_{xx}(\tau) e^{-j2\pi f\tau} d\tau \tag{7}$$

$$P_{yy}(f) = \int_{-\infty}^{+\infty} \rho_{yy}(\tau) e^{-j2\pi f\tau} d\tau \tag{8}$$

$$P_{xy}(f) = \int_{-\infty}^{+\infty} \rho_{xy}(\tau) e^{-j2\pi f\tau} d\tau \tag{9}$$

where $\rho_{xx}(\tau)$ and $\rho_{yy}(\tau)$ are the autocorrelation functions of $x(t)$ and $y(t)$, respectively, and $\rho_{xy}(\tau)$ is the cross-correlation function between $x(t)$ and $y(t)$ (see equation (2)), $f$ is the frequency, $\tau$ is the time lag, and $j$ is the imaginary unit. Therefore, the auto- and cross-spectra quantify how the power of each cumulative spike train, and their shared activity, is distributed across frequency.

In practice, these spectra are typically estimated using Welch's periodogram method (Welch, 1967). In this approach, the cumulative spike trains $x(t)$ and $y(t)$ are segmented into shorter time windows. Each segment is multiplied by a tapering window $w(t)$, typically a 1-s Hanning window, to reduce spectral leakage by attenuating the edges of the segment. The auto- and cross-spectra are then estimated for each segment, and the resulting spectra are averaged across segments to obtain smoother spectral estimates. This segmentation can be performed with or without overlapping between consecutive windows. The duration of the analysis window determines the spectral resolution of the estimate: longer windows provide higher frequency resolution, while shorter windows provide coarser frequency resolution. In addition, because the random splitting of MUs into two cumulative spike trains can influence the coherence estimate, the splitting procedure is repeated for a predefined number of iterations, commonly up to 100 times (Cabral et al., 2025a, Cabral et al., 2024a, Cabral et al., 2024b, Cosentino et al., 2026) (**Figure 5C**). When the number of possible unique group splits is lower than the requested number of iterations, all possible unique splits are used. In each iteration, two new cumulative spike trains are created, and their auto- and cross-spectra are estimated. The auto-spectra and cross-spectrum are then averaged across iterations, resulting in the pooled spectra $\bar{P}_{xx}$, $\bar{P}_{yy}$ and $\bar{P}_{xy}$. The pooled coherence over the iterations is estimated as in Rosenberg et al. (1989):

$$C_{xy}(f) = \frac{\left|\bar{P}_{xy}\right|^2}{\bar{P}_{xx}\bar{P}_{yy}} \tag{10}$$

The coherence $C_{xy}(f)$ ranges from 0 to 1 and quantifies the degree of linear coupling between the two cumulative spike trains at each frequency $f$ (**Figure 5D**). Values close to 0 indicate little or no linear association between the cumulative spike trains at that frequency, whereas values close to 1 indicate strong coupling. In the context of MU analysis, significant coherence between cumulative spike trains is interpreted as evidence of common synaptic input to the motor neuron pool at specific frequencies.

An additional step commonly performed before statistical testing or comparison across conditions is the transformation of coherence values into z-scores. This transformation is particularly useful when comparing coherence estimates obtained across different force levels, experimental conditions, or analyses involving different numbers of MUs (Laine et al., 2015, Del Vecchio et al., 2019, Maillet et al., 2022, Cabral et al., 2025a, Cabral et al., 2024a, Cabral et al., 2024b). The z-score transformation is given by:

$$Zscore(f) = factor \times \tanh^{-1}\left(\sqrt{C_{xy}(f)}\right) \tag{11}$$

where $C_{xy}(f)$ is the pooled coherence at frequency $f$. For non-overlapping windows, the factor is commonly defined as:

$$factor = \sqrt{2L} \tag{12}$$

where $L$ is the number of independent segments used in the spectral estimation. When overlapping windows are used, adjacent segments are not fully independent, and this dependency must be accounted for when calculating the factor. In *openhdemg*, this correction is implemented by estimating an effective number of independent segments based on the overlap between adjacent tapered windows, following the approach proposed by Gallet and Julien (2011). It is also important to note that some studies remove a bias term from the z-score estimate. This bias is often estimated empirically as the average z-score coherence between 250 and 500 Hz, because this frequency band is expected to contain no significant physiological coherence between MU cumulative spike trains (Baker et al., 2003, Castronovo et al., 2015).

Finally, coherence is commonly summarized within predefined frequency bands. In the MU literature, this is typically done by calculating either the average or the area under curve of the coherence or z-score coherence within frequency bands such as delta (1-5 Hz), alpha (5-15 Hz), and beta (15-35 Hz). These band-specific estimates allow the user to quantify the strength of common synaptic input within physiologically relevant frequency ranges and to compare common input across muscles, tasks, experimental conditions, or parameter choices.

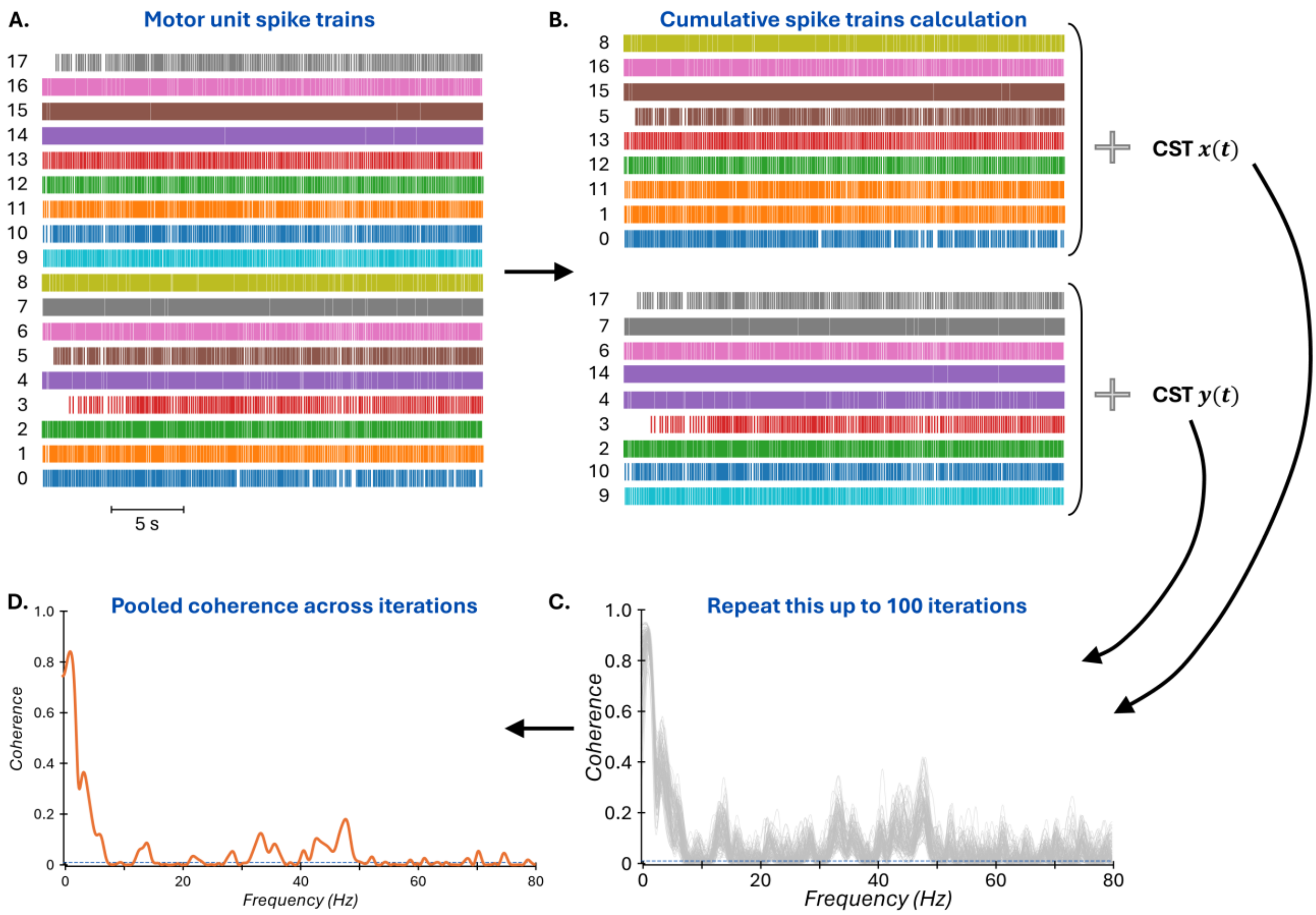

***Figure 5: Step-by-step estimation of the pooled coherence from motor unit spike trains***. *(A) Motor unit spike trains obtained from decomposed high-density surface electromyography (HDsEMG) signals. (B) The population of motor unit spike trains is randomly split into two non-overlapping groups of equal size. Within each group, the binary spike trains are summed to obtain two cumulative spike trains (CSTs). (C) The random splitting procedure is repeated across multiple iterations, in this example 100 iterations, generating different pairs of CSTs from the same motor unit population. For each iteration, auto- and cross-spectra are estimated from the two CSTs using Welch's method. (D) Pooled intramuscular coherence is obtained by averaging the spectral estimates across iterations and calculating coherence between the two CST groups. The resulting coherence curve quantifies the frequency-specific coupling between motor unit populations and is interpreted as an estimate of oscillatory common synaptic input to the motor neuron pool.*

### *4.1.3. Statistical significance*

As for the common drive index, a confidence level can be calculated to determine which portions of the coherence or z-score coherence curves exceed the values expected by chance. This threshold is particularly important when coherence is summarized within predefined frequency bands, because the average coherence or the area under the curve is commonly calculated only from the portions of the curve that exceed the confidence level. Therefore, the confidence level determines which frequency bins are considered to represent significant common oscillatory input to the motor neuron pool. Two main approaches can be used to define the confidence level. The first possibility is to use an empirical high-frequency bias, calculated as the average coherence or z-score coherence within a frequency range in which significant physiological coherence is not expected, for example, between 250 and 500 Hz (Baker et al., 2003, Castronovo et al., 2015). This approach provides a data-driven estimate of the baseline level of coherence and can be useful to account for residual bias in the spectral estimate. In this case, only coherence or z-score coherence values above the high-frequency bias are considered when computing band-specific estimates. Alternatively, a theoretical confidence level can be calculated from the expected distribution of coherence under the null hypothesis of no linear association between the two cumulative spike trains. For raw pooled coherence, the 95% confidence level is commonly defined as in Rosenberg et al. (1989):

$$CL = 1 - 0.05^{1/(L-1)} \tag{13}$$

where L is the number of independent segments used in Welch's spectral estimation. For z-score coherence, the confidence level is commonly defined as 1.65, which corresponds to the 95$^{th}$ percentile of the standard normal distribution. Therefore, z-score coherence values above 1.65 are considered significant at the 5% level. Both approaches are implemented in *openhdemg* and, in both cases, the confidence level is used to threshold the coherence or z-score coherence curve before calculating band-specific metrics, such as the average coherence or the area under the curve in the delta, alpha, beta and gamma frequency bands.

### *4.1.4. Parameter sensitivity and recommendations*

The estimation of the common drive index is implemented in *openhdemg* through the function *pooled_intramuscular_coherence*, with the main structure described in the Supplementary Material. For pooled coherence, we systematically tested how three main parameters affected the estimates (**Figure 6**): the number of MUs included in each cumulative spike train, the duration of the Hanning window used in Welch's spectral estimation, and the overlap between consecutive Welch windows. These parameters influence the robustness, spectral resolution, and variability of the coherence estimate and should therefore be carefully considered when interpreting frequency-domain estimates of common synaptic input.

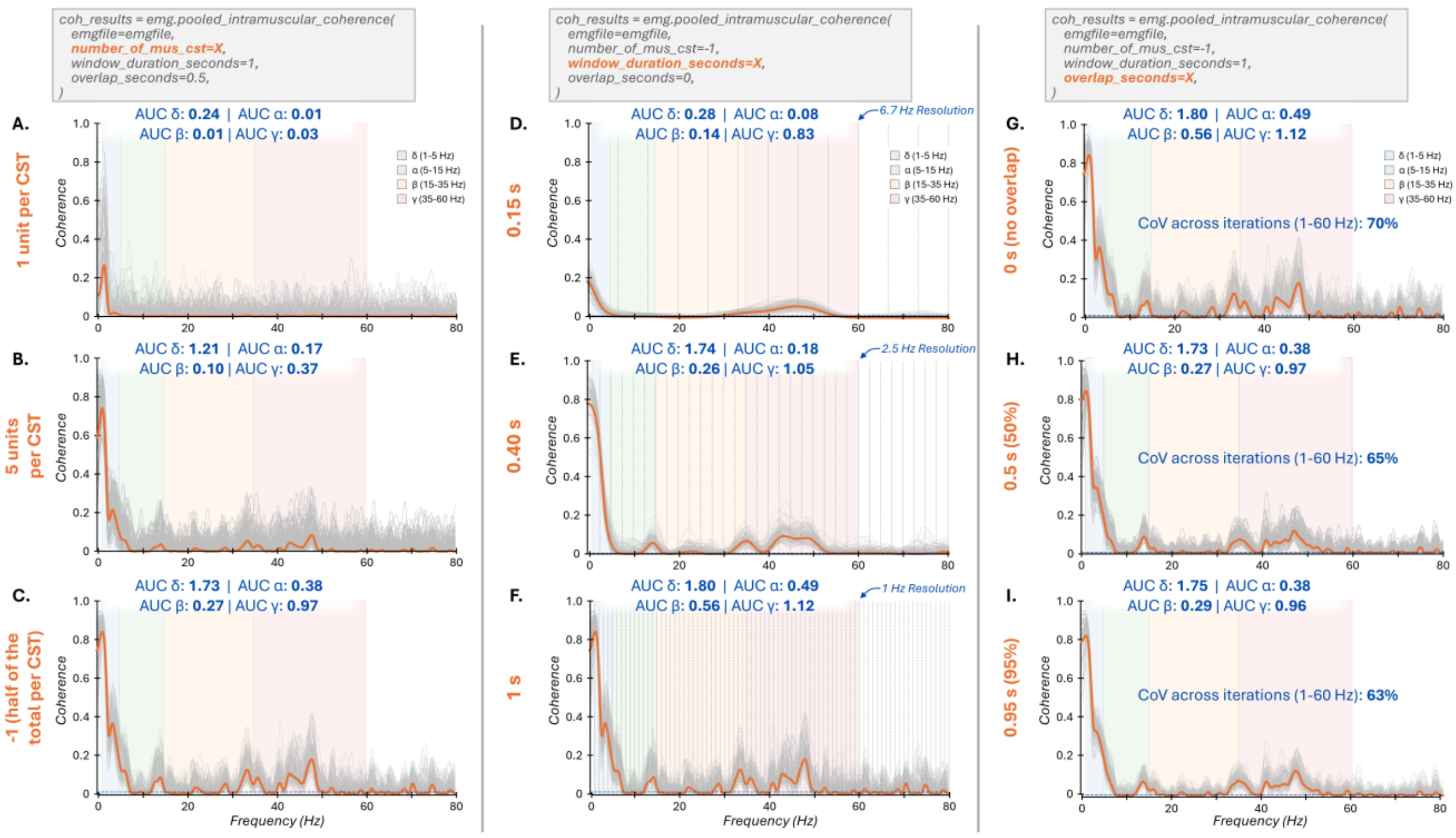

***Figure 6: Parameter sensitivity of pooled coherence.*** *Pooled coherence was estimated from cumulative spike trains obtained by randomly splitting the population of motor unit spike trains into two non-overlapping groups. Panels A–C show the effect of changing the number of motor units included in each cumulative spike train: 1 motor unit per cumulative spike train (A), 5 motor units per cumulative spike train (B), and half of the total number of available motor units per cumulative spike train, corresponding to 9 motor units in this dataset (C). Note that the area under the curve increased from A to C in the delta (from 0.24 to 1.73), alpha (from 0.01 to 0.38), beta (from 0.01 to 0.27), and gamma (from 0.03 to 0.97) bands. Panels D–F show the effect of changing the duration of the Hanning window used in Welch's spectral estimation: 0.15 s (D), 0.4 s (E), and 1 s (F), corresponding to frequency resolutions of approximately 6.7 Hz, 2.5 Hz, and 1 Hz, respectively. As the Hanning window duration increased, the pooled coherence estimates increased, and the spectral profile became more detailed due to improved frequency resolution. These differences were reflected in the band-specific estimates, in which delta (from 0.28 to 0.18), alpha (from 0.08 to 0.49), beta (from 0.14 to 0.56), and gamma (from 0.83*

*to 1.12) bands increased from 0.15-s window to 1-s window. Panels G–I show the effect of changing the overlap between consecutive Welch windows while using a 1-s Hanning window: 0 s overlap (G), 0.5 s overlap (H), and 0.95 s overlap (I). Gray lines represent coherence curves obtained from individual random motor unit splits, whereas the orange line represents the pooled coherence estimate. Increasing the overlap between Welch windows reduced the variability of the coherence estimates across random splits. This effect was reflected in the band-specific estimates, with the area under the curve decreasing from G to I in the delta (from 1.80 to 1.75), alpha (from 0.49 to 0.38), beta (from 0.56 to 0.29), and gamma (from 1.12 to 0.96) bands. AUC, area under the curve; CoV, coefficient of variation.*

First, we tested the effect of the number of MUs included in each cumulative spike train. To do this, we calculated coherence using 1 MU per cumulative spike train (**Figure 6A**), 5 MUs per cumulative spike train (**Figure 6B**), and half of the total number of available MUs per cumulative spike train (**Figure 6C**). In this dataset, 18 MUs were available; therefore, the last condition corresponded to 9 MUs per cumulative spike train. As expected (Negro and Farina, 2012, Negro et al., 2016b), increasing the number of MUs included in each cumulative spike train increased the magnitude of the coherence curve. This effect was reflected in the band-specific estimates, with the area under the curve increasing from **Figure 6A** to **Figure 6C** in the delta (from 0.24 to 1.73), alpha (from 0.01 to 0.38), beta (from 0.01 to 0.27), and gamma (from 0.03 to 0.97) bands. This behavior is consistent with the rationale of pooled coherence: summing the spike trains of multiple MUs increases the effective sampling of the common synaptic input to the motor neuron pool and attenuates the influence of independent discharge variability. Therefore, coherence estimates obtained from cumulative spike trains containing larger MU samples are generally more robust than estimates obtained from individual MU pairs.

Second, we tested the effect of the duration of the Hanning window used in Welch's spectral estimation. We compared window durations of 0.15 s (**Figure 6D**), 0.4 s (**Figure 6E**), and 1 s (**Figure 6F**), corresponding to spectral resolutions of approximately 6.7 Hz, 2.5 Hz, and 1 Hz, respectively. With the 0.15-s window, the coherence curve showed low values and coarse spectral resolution, making it difficult to determine coherence within narrow frequency bandwidths, particularly at low frequencies. As the Hanning window duration increased, the pooled coherence estimates increased, and the spectral profile became more detailed due to improved frequency resolution. These differences were reflected in the band-specific estimates, in which delta (from 0.28 to 0.18), alpha (from 0.08 to 0.49), beta (from 0.14 to 0.56), and gamma (from 0.83 to 1.12) bands increased from 0.15-s window to 1-s window. Therefore, the choice of window duration depends on the objective of the analysis. Shorter windows may be useful when temporal localization is prioritized, but they provide poor frequency resolution and may obscure narrow-band coherence. Longer windows improve frequency resolution and are more appropriate when the goal is to quantify coherence within specific frequency bands. A 1-s Hanning window provides a frequency resolution of approximately 1 Hz and represents a practical compromise that has been widely used in the MU literature.

Finally, we tested the effect of overlapping between consecutive Welch windows. For this analysis, the duration of the Hanning window was fixed at 1 s, and overlap was varied from 0 s (non-overlapping; **Figure 6G**), to 0.5 s (50% overlap; **Figure 6H**), and 0.95 s (95% overlap; **Figure 6I**). Increasing the overlap reduced the variability of the coherence estimates across iterations, as visually indicated by the smaller dispersion of the gray curves representing the coherence curves obtained from each random split of MUs. To quantify this effect, we calculated the coefficient of variation across the iteration-wise coherence curves up to 80 Hz, which decreased from 70% with no overlap to 63% with 95% overlap. This effect was reflected in the band-specific estimates, with the area under the curve decreasing from **Figure 6G** to **Figure 6I** in the delta (from 1.80 to 1.75), alpha (from 0.49 to 0.38), beta (from 0.56 to 0.29), and gamma (from 1.12 to 0.96) bands. Therefore, overlapping windows can provide more stable and less variable spectral estimates. However, adjacent overlapping segments are not statistically independent, and this dependency should be considered when calculating z-score transformation, as detailed in the previous subsection. Thus, although high overlap can improve the empirical stability of the estimate, it should not be interpreted as a proportional increase in independent information.

Based on these observations, it is recommended to use half of the total number of available MUs in each cumulative spike train whenever possible, while ensuring that the two cumulative spike trains are composed of non-overlapping MU groups. This maximizes the population-level representation of common synaptic input while preserving independent groups for coherence estimation. It is important to note that, as demonstrated in **Figure 6A-C**, increasing the number of MUs contributing to the cumulative spike trains systematically increases the magnitude of the coherence estimates across frequency bands. Therefore, when comparing coherence between experimental conditions, interventions, muscles, populations, or time points, it is important to standardize the number of MUs included in the analysis. Otherwise, apparent differences in coherence may simply reflect differences in the number of MUs contributing to the cumulative spike trains rather than true physiological differences in common synaptic input (Castronovo et al., 2015, Negro and Farina, 2012). For spectral estimation, a 1-s Hanning window is a practical default because it provides approximately 1-Hz frequency resolution and allows the identification of coherence within commonly analyzed frequency bands. Finally, 50% overlap can be used as a conservative default for Welch's method, although higher overlap values may also be used to obtain smoother and less variable estimates, as performed in recent studies from our group (Cabral et al., 2025a, Cabral et al., 2024a, Cabral et al., 2024b). When high overlap is used, the effective number of independent segments should be considered for statistical testing and z-score transformation. In all cases, the number of MUs per cumulative spike train, window duration, overlap, frequency resolution, and confidence-level procedure should be explicitly reported.

## 4.2. Proportion of common input index

### *4.2.1. Physiological interpretation*

The proportion of common input (PCI) index is based on the approach proposed by Negro et al. (2016b). This method was developed to overcome an important limitation of classical pairwise correlation and coherence analyses, which can indicate whether MU spike trains share common fluctuations but do not directly quantify the proportion of common input relative to the overall synaptic input received by the motor neurons. The method relies on the principle that the cumulative spike train progressively enhances the representation of common synaptic input as more MUs are included, while independent components of the synaptic input are attenuated through population averaging (Negro and Farina, 2012, Negro et al., 2016b). Therefore, by estimating coherence between cumulative spike trains composed of increasing numbers of MUs, it is possible to model how shared input becomes more strongly represented in the motor neuron pool output. The experimental coherence values obtained for different cumulative spike train sizes are then fitted with a nonlinear model that separates the contribution of common and independent input components. The resulting PCI index obtained from parameters of this fit provides an estimate of the relative contribution of common synaptic input to the total synaptic input received by the motor neuron pool. In its original formulation, the method was applied to low-frequency common input below 5 Hz (delta band), reflecting the slow-varying components of synaptic input that are most strongly transmitted to force output during isometric contractions (Farina et al., 2014, Negro et al., 2009, Cabral et al., 2025b, Thompson et al., 2018).

*4.2.2. Step-by-step estimation*

The initial steps for estimating the proportion of common input index are similar to those described for pooled coherence. The analysis starts from the binary spike trains of the detected MUs, which are used to form two non-overlapping cumulative spike trains. For each cumulative spike train, the binary spike trains of the selected MUs are summed, and coherence between the two cumulative spike trains is estimated as described in Section 4.1.2 (**Figure 5**).

The main difference from pooled coherence is that the coherence estimation is repeated while systematically varying the number of MUs included in each cumulative spike train. Let $N$ be the total number of detected MUs. The number of MUs per cumulative spike train, $n$, is varied from 1 to the maximum possible value, corresponding to half of the total number of detected MUs. For each value of $n$, the MUs are randomly divided into two non-overlapping groups containing $n$ MUs each. Coherence is then calculated between the two cumulative spike trains obtained from these groups. This procedure is repeated for multiple iterations, commonly up to 100 times, and the pooled coherence is calculated for each value of $n$ (**Figure 7A**). When the number of possible unique group splits is lower than the requested number of iterations, all possible unique splits are used. After obtaining the pooled coherence curve for each cumulative spike train size, the average coherence is calculated within predefined frequency bands. In the original formulation, Negro et al. (2016b) focused on the low-frequency band below 5 Hz, reflecting the slow-varying common synaptic input to the motor neuron pool. In the present implementation, the same approach was also extended to other frequency bands, including alpha (5-15 Hz) and beta (15-35 Hz). For each frequency band, this procedure results in a curve describing how the average coherence changes as a function of the number of MUs included in each cumulative spike train (**Figure 7B**). The relationship between the number of MUs per cumulative spike train and the average coherence is then fitted using the model proposed by Negro et al. (2016b):

$$C(n) = \frac{(n^2A)^2}{(nB + n^2A)^2} \tag{13}$$

where $C(n)$ is the average coherence for the selected bandwidth obtained when each cumulative spike train contains $n$ MUs, and $A$ and $B$ are model parameters estimated by least-squares fitting. In this framework, $A$ is associated with the common component of the synaptic input, whereas $B$ is associated with the independent component. Therefore, the fitted curve describes how the representation of common synaptic input in the cumulative spike trains increases as more MUs are included in the estimate (**Figure 7B**). The proportion of common input index is calculated from the fitted parameters as:

$$PCI = \sqrt{A/B}$$

Therefore, higher PCI values indicate that a greater proportion of the synaptic input represented in the motor neuron pool output is common across MUs. When the analysis is applied separately to different frequency bands, the PCI can be interpreted as a band-specific estimate of the proportion of common input. However, although the PCI framework can be extended to higher frequency bands, the resulting estimates may be more susceptible to distortion and estimation variability than those obtained in the low-frequency range originally proposed (Negro et al., 2016b).

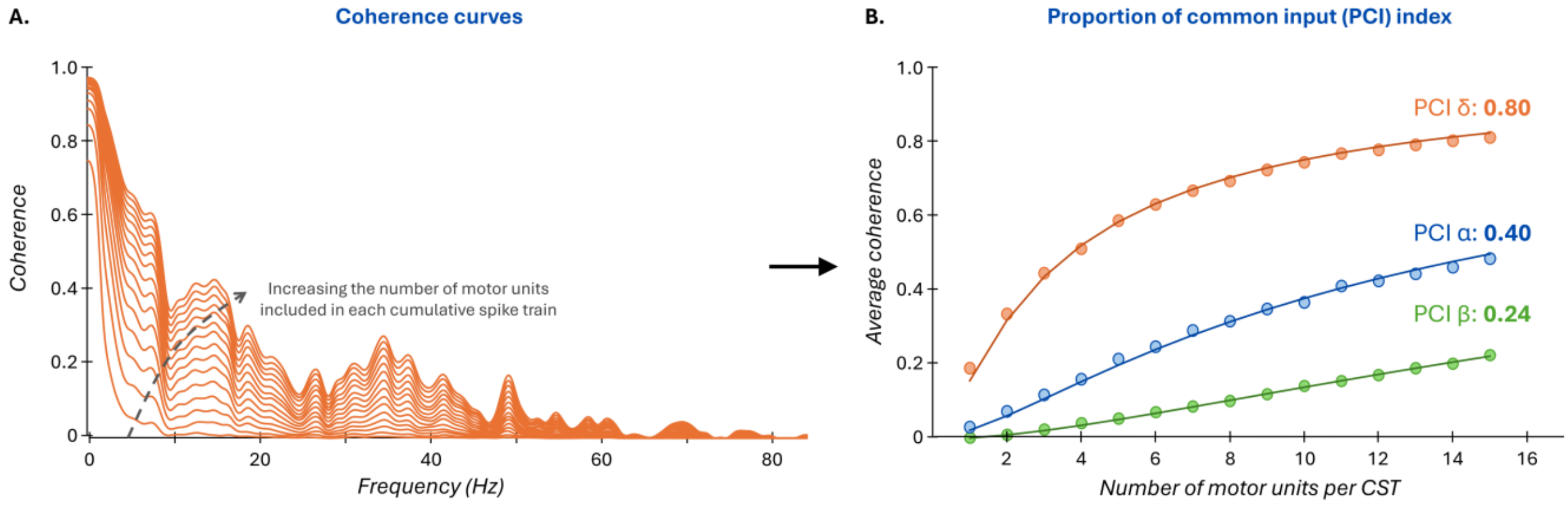

***Figure 7: Step-by-step estimation of the proportion of common input (PCI) index from motor unit spike trains.*** *(A) Coherence curves obtained by progressively increasing the number of motor units included in each cumulative spike train. In this example, 31 motor units were identified; therefore, the number of motor units per cumulative spike train was varied from 1 to 15. As more motor units were included in each cumulative spike train, the coherence magnitude increased, reflecting the enhanced representation of common synaptic input in the population spike trains. (B) Relationship between the number of motor units per cumulative spike train and the average coherence calculated within predefined frequency bands. Circles represent the average coherence values for the delta, alpha, and beta bands, whereas solid lines represent the nonlinear least-squares fits used to estimate the proportion of common input index. The PCI is calculated from the fitted model parameters and provides a band-specific estimate of the relative contribution of common synaptic input with respect to the total synaptic input.*

### *4.2.4. Parameter sensitivity and recommendations*

The estimation of the PCI index is implemented in *openhdemg* through the function *pci_index*, with the main structure described in the Supplementary Material. For this approach, we systematically tested how three main parameters affected the estimates (**Figure 8**): the total number of MUs included in the analysis, the duration of the Hanning window used in Welch's spectral estimation, and the overlap between consecutive Welch windows.

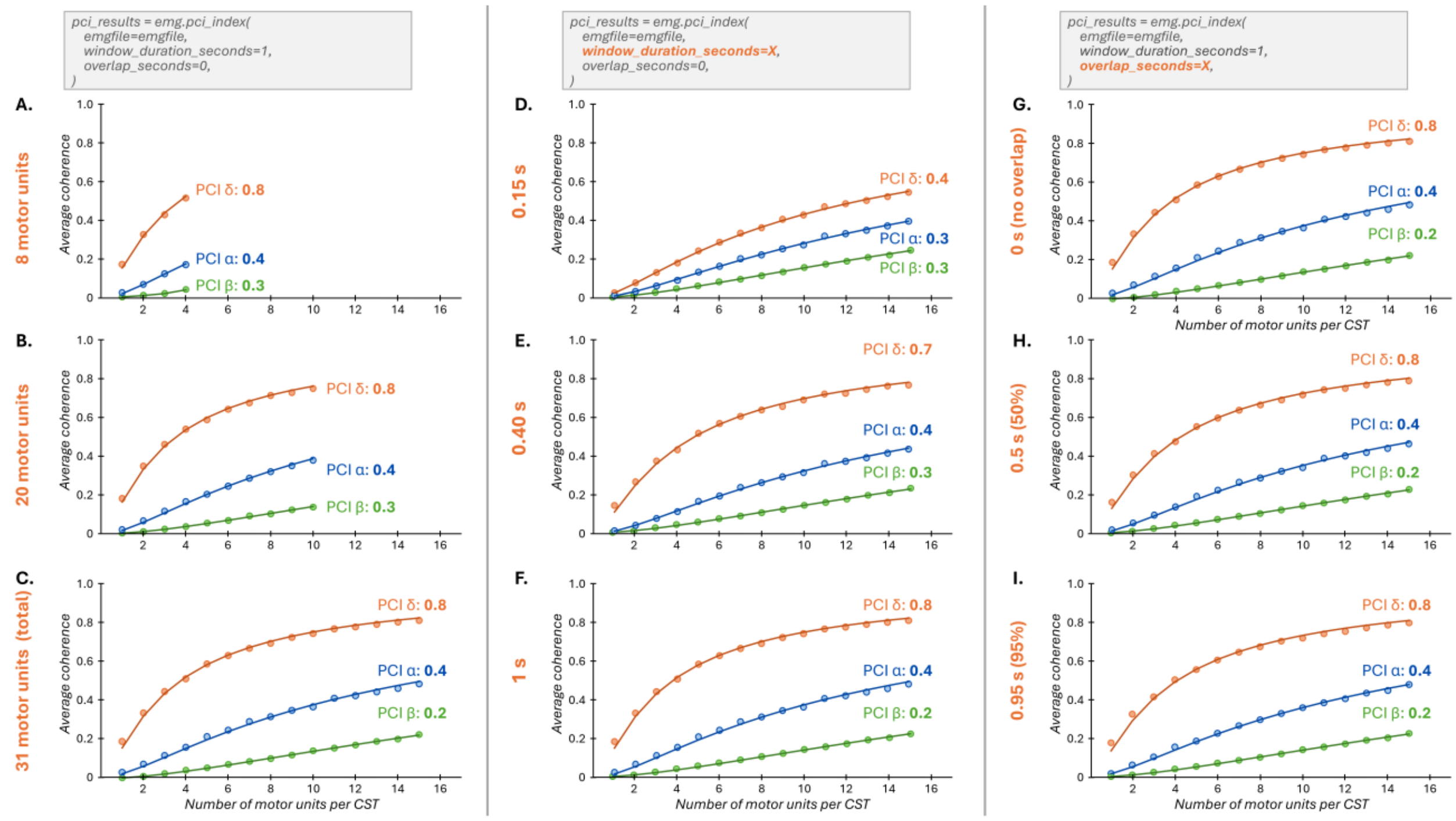

***Figure 8: Parameter sensitivity of the proportion of common input index (PCI)****. The PCI was estimated by fitting the relationship between the average coherence and the number of motor units included in each cumulative spike train (CST). Panels A–C show the effect of changing the total number of motor units available for the analysis: 8 motor units (A), 20 motor units (B), and 31 motor units (C). Increasing the number of motor units resulted in a clearer saturation of the coherence-versus-CST-size relationship, although the resulting PCI estimates were relatively similar across conditions. Panels D–F show the effect of changing the duration of the Hanning window used in Welch's spectral estimation: 0.15 s (D), 0.4 s (E), and 1 s (F). Longer windows increased the estimated PCI in the delta band by improving the resolution of low-frequency coherence, whereas alpha- and beta-band PCI estimates were less affected. Panels G–I show the effect of changing the overlap between consecutive Welch windows while using a 1-s Hanning window: 0 s overlap (G), 0.5 s overlap (H), and 0.95 s overlap (I). Changes in overlap had minimal influence on the resulting PCI estimates across all frequency bands. In each panel, circles represent the average coherence values obtained for different cumulative spike train sizes, whereas solid lines represent the nonlinear least-squares fit used to estimate the PCI. Delta-, alpha-, and beta-band PCI values are reported in orange, blue, and green, respectively.*

First, we tested the effect of the number of MUs included in the analysis. To investigate this effect, we randomly selected 8 MUs (allowing up to 4 MUs per cumulative spike train; **Figure 8A**), 20 MUs (allowing up to 10 MUs per cumulative spike train; **Figure 8B**), and used the full population of 31 MUs available in the dataset (allowing up to 15 MUs per cumulative spike train; **Figure 8C**). As the number of available MUs increased, the relationship between the number of MUs in the cumulative spike train and the average coherence became progressively less steep and showed a clearer saturation behavior at larger cumulative spike train sizes. This behavior is expected because larger MU populations provide a more complete representation of the common synaptic input. In contrast, when only a few MUs are available, the fitted curve may become steeper, potentially leading to overestimation of the PCI index. Interestingly, despite these differences in curve shape, the PCI estimates obtained with 8 MUs were already very similar to those obtained with 20 or 31 MUs. Therefore, although larger MU populations provide more robust estimates, meaningful PCI estimates can still be obtained from relatively small MU samples.

Second, we tested the effect of the duration of the Hanning window used in Welch's spectral estimation. As for pooled coherence, we compared window durations of 0.15 s (**Figure 8D**), 0.4 s (**Figure 8E**), and 1 s (**Figure 8F**). The PCI estimated in the delta band increased substantially, from 0.4 with a 0.15-s window to 0.8 with a 1-s window. This behavior is a direct consequence of the effects observed in **Figures 6D-F** for pooled coherence. Longer analysis windows improve frequency resolution and allow a more accurate estimation of low-frequency coherence, which is the primary frequency range targeted by the original PCI formulation. In contrast, shorter windows provide poorer frequency resolution and tend to underestimate low-frequency coherence, leading to lower PCI values.

Finally, we tested the effect of overlapping between consecutive windows. Using a fixed Hanning window duration of 1 s, overlap was varied from 0 s (no-overlapping; **Figure 8G**) to 0.5 s (50% overlap; **Figure 8H**), and 0.95 s (95% overlap; **Figure 8I**). In contrast to pooled coherence, changing the overlap produced only minor changes in the PCI estimates across all frequency bands. This result is expected because overlap mainly affects the variability and smoothness of the coherence estimates rather than their average magnitude. Consequently, the fitted curves and the resulting PCI estimates remained largely unchanged across overlap conditions.

Based on these observations, all available cleaned MUs should be included whenever possible, as larger MU populations provide a more representative characterization of the common synaptic input. Nevertheless, the present results suggest that relatively few MUs may already provide reliable PCI estimates, particularly when saturation of the fitted curve is observed. A 1-s Hanning window appears appropriate, consistent with the original proposition of (Negro et al., 2016b), because it provides adequate frequency resolution for estimating low-frequency coherence and yields stable PCI estimates. Finally, no overlap between windows can be used, following the original implementation of the method, although overlap had little influence on the resulting PCI estimates in the present analysis. Importantly, because the original theoretical framework was developed to quantify the slow-varying common synaptic input represented in the delta band, and because higher-frequency oscillations are increasingly affected by the nonlinear transformation from synaptic input to motor neuron output spike trains, it is strongly recommended to interpret the delta-band PCI as the primary physiological estimate of the proportion of common synaptic input. Estimates obtained in higher-frequency bands should be considered exploratory and interpreted with appropriate caution.

## 5. Network-information nonlinear framework

### *5.1. Physiological interpretation*

Most approaches for estimating common synaptic input from MU spike trains, including the time- and frequency-domain methods described in the previous sections, rely predominantly on linear techniques. These methods have provided important insights into the shared input received by motor neuron populations, but they impose assumptions that may not fully capture nonlinear dependencies among MU outputs. Moreover, even when population-level signals such as cumulative spike trains are used, the analysis often remains based on pairwise comparisons between two signals. As a result, these approaches quantify the magnitude or spectral content of shared fluctuations but provide limited information about how dependencies are organized across the MU population as a functional network. To characterize potential nonlinear interactions among MU discharge patterns, a recently developed framework combining information and network theories (O'Reilly and Delis, 2024, O'Reilly and Delis, 2022) can be applied to MU activity (Cabral et al., 2025b). In this framework, mutual information is used to quantify the statistical dependence between pairs of MU discharge rates. Mutual information provides a general measure of dependence between two random variables because it does not assume that the relationship is linear or monotonic (Cover, 1999). Therefore, in contrast to correlation-based approaches, mutual information can capture nonlinear dependencies that may be present in MU activity. When applied to smoothed MU discharge rates, this approach estimates the degree of shared information between each pair of MUs and represents these dependencies as a weighted network (Cabral et al., 2025b). After thresholding the network to retain only relevant dependencies, graph-theoretical tools can be used to identify groups of MUs that are more strongly connected to each other than to the rest of the population. These groups can be interpreted as functional MU communities, reflecting subsets of MUs with similar or strongly coupled discharge rate fluctuations. In this network, each node corresponds to one MU, and each edge represents the strength of the mutual information between two MUs. Network theory has also been used recently to represent the linear dependencies (i.e., pairwise cross-correlations) between MUs (Hug et al., 2023b).

*5.2. Step-by-step estimation*

The first preprocessing steps are similar to those used for the time-domain approaches described previously. The binary spike train of each MU (**Figure 9A**) is converted into a smoothed discharge rate using a Hanning window, edge portions affected by convolution and filtering are removed, and slow offsets or trends are attenuated using a zero-phase high-pass filter (**Figure 9B**). This results in a matrix of detrended smoothed discharge rates, with dimensions $T\ x\ N$, where $T$ is the number of time samples, and $N$ is the number of MUs.

The detailed mathematical formulation of the information- and network-theoretic procedures has been described extensively elsewhere (Ince et al., 2017, O'Reilly and Delis, 2024, O'Reilly and Delis, 2022).

Therefore, only the main processing steps are summarized here. First, each smoothed discharge rate is transformed using copula normalization (Ince et al., 2017). This procedure ranks the samples of each MU discharge-rate signal, rescales the ranks to empirical cumulative probabilities, and then maps these probabilities onto a standard Gaussian distribution. This transformation preserves the rank-based dependence structure between variables while removing the influence of their marginal distributions. Mutual information between two smoothed discharge rates $y_i(t)$ and $y_j(t)$ of the $i^{th}$ and $j^{th}$ MUs can be expressed as:

$$MI(Y_i, Y_j) = H(Y_i) + H(Y_j) - H(Y_i, Y_j)$$

where $Y_i$ and $Y_j$ denote the samples of the smoothed discharge rate signals over time, $H(Y_i)$ and $H(Y_j)$ are their marginal entropies, and $H(Y_i, Y_j)$ their joint entropy. Thus, mutual information quantifies the reduction in uncertainty about one MU discharge rate when the other is known. After copula normalization, mutual information was estimated using a Gaussian copula formulation. For two scalar copula-normalized MU discharge rates, this can be written as:

$$MI(\tilde{Y}_i, \tilde{Y}_j) = -\frac{1}{2}\log_2(1 - {\rho_{ij}}^2)$$

where $\tilde{Y}_i$ and $\tilde{Y}_j$ are the copula-normalized MU discharge rates, and $\rho_{ij}$ is the Pearson correlation coefficient between them. This formulation provides an efficient estimate of statistical dependence while reducing the influence of the marginal distributions of individual MU discharge rates via an initial rank-based copula transformation.

Gaussian copula mutual information is calculated between all possible pairs of MUs. Conceptually, this step is similar to the pairwise structure used in the common drive index, but instead of calculating cross-correlation between MU discharge rates, the method calculates the amount of shared information between them. The result is a symmetric pairwise mutual information matrix of dimension $N\ x\ N$ MUs, in which each element represents the strength of the statistical dependence between two MUs (**Figure 9B**). To reduce the influence of weak or potentially spurious dependencies, the weighted network is thresholded using a modified percolation analysis (Gallos et al., 2012). This procedure progressively removes weak connections until the global connected structure of the network begins to be affected, providing a data-driven threshold for retaining the most relevant dependencies.

Finally, the thresholded network is analyzed using graph-theoretical tools (**Figure 9D**). Significant mutual information values define the edges of the MU network, and community-detection procedures are used to identify groups of strongly connected MUs. In the present implementation, link communities are identified using hierarchical clustering of network edges (Ahn et al., 2010), allowing MUs to participate in multiple functional groups when their connections support such an organization. The resulting communities provide a low-dimensional description of the MU population based on its nonlinear dependency structure.

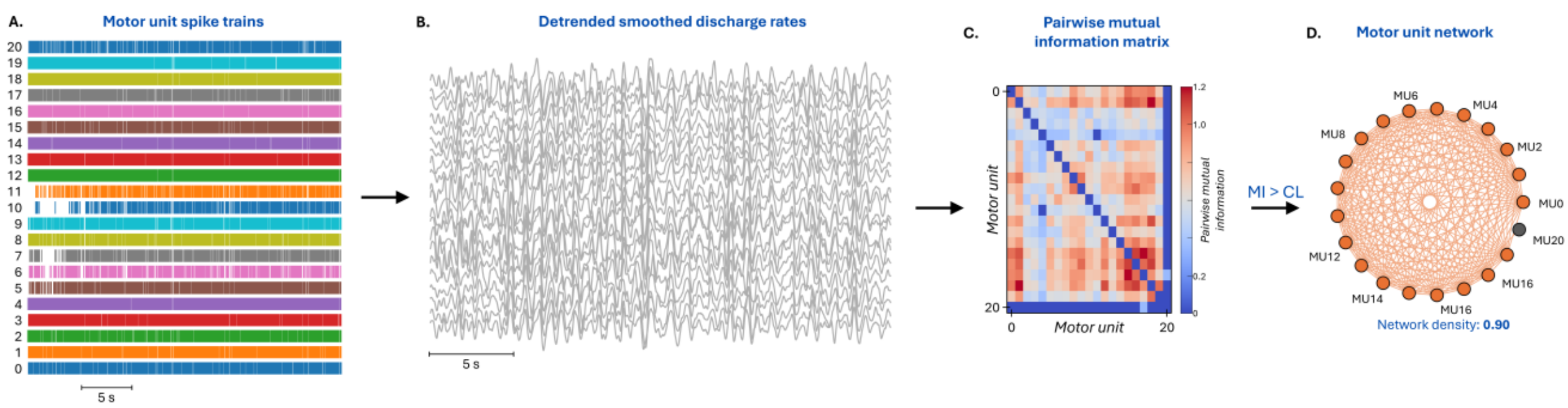

***Figure 9:*** *Step-by-step estimation of the network-information nonlinear framework from motor unit spike trains.*

## *5.3. Statistical significance*

To identify the most relevant pairwise dependencies among MUs, the mutual information matrix was thresholded using a modified percolation analysis adapted from Gallos et al. (2012). In this procedure, the weighted network is progressively thresholded by removing weaker edges, and the connectivity structure of the graph is monitored as the threshold increases. The selected threshold corresponds to the highest mutual information value that preserves the global connected structure of the network before it fragments into disconnected components. All mutual information values below this threshold are set to zero, whereas the remaining edges are retained as relevant functional dependencies between MUs. This procedure provides a data-driven criterion to remove weak or potentially spurious connections while preserving the main topology of the MU network. The resulting thresholded matrix is then used as the adjacency matrix for graph visualization and community detection.

### *5.4. Parameter sensitivity and recommendations*

The estimation of the network-information framework using mutual information is implemented in *openhdemg* through the function *smoothed_dr_mutualinformation*, with the main structure described in the Supplementary Material. For this approach, we systematically tested how two main parameters affected the estimates (**Figure 10**): the total number of MUs included in the analysis and the duration of the Hanning window used to smooth the MU discharge rates.

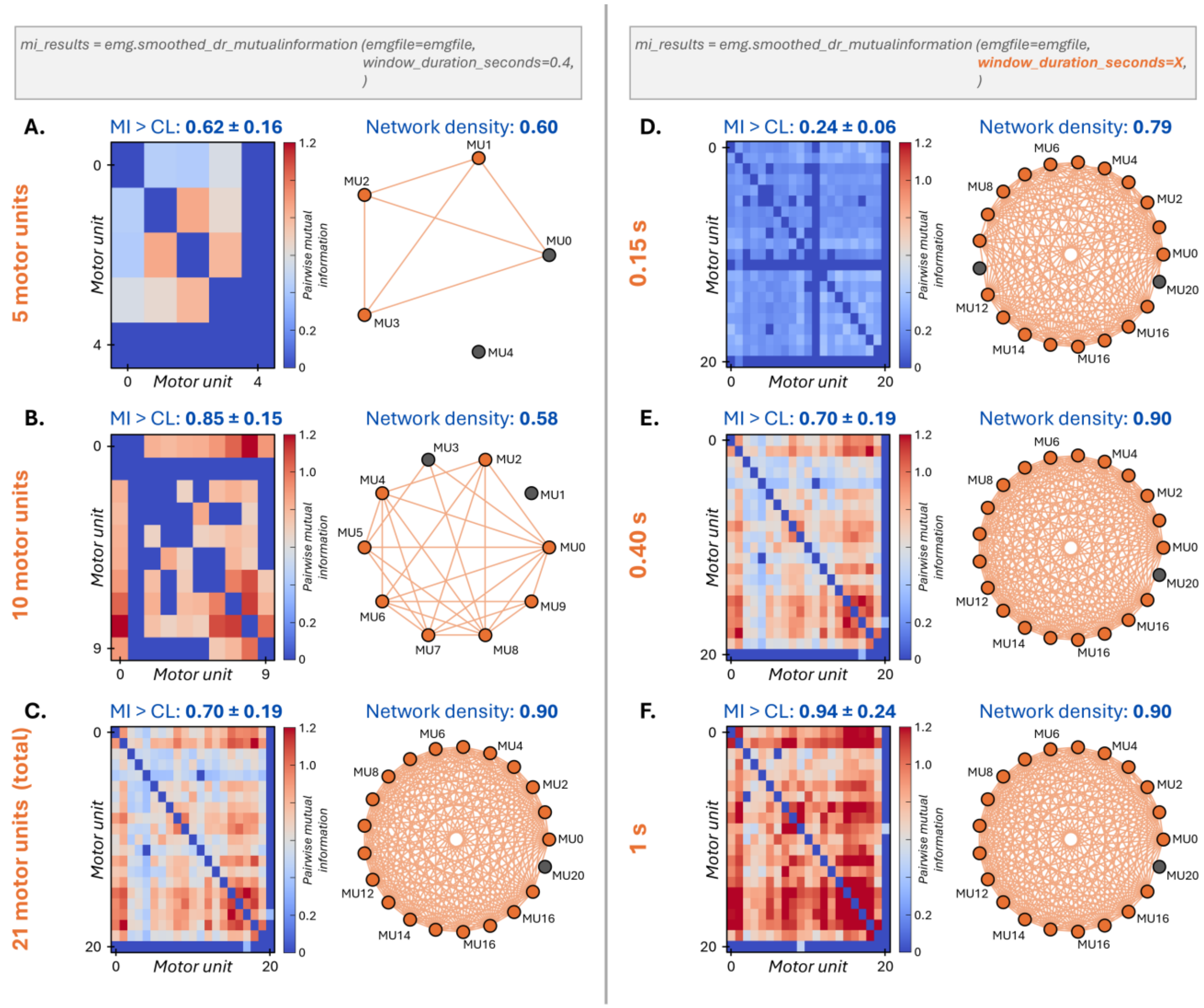

***Figure 10: Parameter sensitivity of the network-information framework based on mutual information.*** *The network-information framework was applied to detrended smoothed motor unit discharge rates to estimate nonlinear pairwise dependencies among motor units and characterize their network organization. Panels A–C show the effect of changing the number of motor units included in the analysis: 5 motor units (A), 10 motor units (B), and all 21 motor units available in the dataset (C). Increasing the number of motor units resulted in progressively denser mutual information networks, as shown by the reduction in dark-blue regions in the pairwise mutual information matrices and by the increase in retained network connections. Network density, defined as the proportion of retained connections relative to the total number of possible connections, increased from 0.60 with 5 motor units to 0.90 with 21 motor units. Panels D–F show the effect of changing the duration of the Hanning window used to smooth the binary spike trains before mutual information estimation: 0.15 s (D), 0.4 s (E), and 1 s (F). Longer smoothing windows increased the average thresholded mutual information across motor unit pairs and resulted in denser networks, reflecting the greater emphasis on slower discharge-rate fluctuations shared across motor units. In each panel, the matrix represents pairwise mutual information values between motor units, whereas the network representation shows motor units as nodes and retained mutual information values as edges. Orange nodes indicate motor units assigned to the first low-dimensional component identified using link-community analysis, whereas gray nodes indicate motor units not assigned to this component.*

First, we tested the effect of the number of MUs included in the analysis. To investigate this effect, we randomly selected 5 MUs (**Figure 10A**), 10 MUs (**Figure 10B**), and used the full population of 21 MUs available in the dataset (**Figure 10C**). Increasing the number of MUs resulted in progressively denser mutual information networks. This can be observed both in the pairwise mutual information matrices, where fewer dark-blue regions (i.e., pairwise mutual information below the threshold) were present as the number of MUs increased, and in the corresponding network representations. Network density, defined as the proportion of retained connections relative to the total number of possible connections, increased from 0.60 when 5 MUs were included to 0.90 when all 21 MUs were used. These results indicate that larger MU samples provide a more complete representation of the nonlinear dependency structure within the MU population. In addition, the number of MUs represented by orange nodes, indicating MUs participating in the first low-dimensional component identified by link-community analysis, remained high across conditions, whereas gray nodes represented MUs not assigned to this component.

Second, we tested the effect of the duration of the Hanning window used to smooth the binary spike trains before mutual information estimation. As for the previous time-domain approaches, we compared smoothing windows of 0.15 s (**Figure 10D**), 0.4 s (**Figure 10E)**, and 1 s (**Figure 10F**). Increasing the smoothing duration substantially increased the strength of the detected dependencies among MUs. The average thresholded mutual information across MU pairs increased from 0.24 ± 0.06 with the 0.15-s window to 0.94 ± 0.24 with the 1-s window. Network density also increased, from 0.79 to 0.90. This behavior is expected because longer smoothing windows attenuate high-frequency discharge variability and emphasize slower fluctuations that are more likely to be shared across MUs. However, as discussed for the common drive index and principal component analysis, higher mutual information values or denser networks should not automatically be interpreted as stronger common synaptic input. Longer smoothing windows progressively restrict the analysis to lower-frequency fluctuations and reduce the temporal resolution of the estimated dependency structure.

Based on these observations, using as many well-identified and sufficiently cleaned MUs as available may provide a more representative characterization of the network structure underlying MU discharge-rate fluctuations. However, comparisons across conditions, muscles, populations, or time points should consider differences in MU yield, as the number of MUs included can influence the resulting network density and dependency structure. For smoothing, a 0.4-s Hanning window is a practical default, consistent with the time-domain approaches described previously. This window duration provides a reasonable compromise between reducing high-frequency discharge variability and preserving the temporal structure of shared fluctuations among MUs. In all cases, the number of MUs, smoothing window duration, thresholding procedure, network density, and community-detection approach should be explicitly reported.

## 6. From HDsEMG decomposition to motor unit cleaning and common synaptic input estimation

Because all estimates of common input strongly depend on the quality of the decomposed MU spike trains, MU decomposition, inspection, and cleaning should not be considered simple preprocessing steps. Rather, they represent essential stages of the analysis pipeline, directly influencing the physiological validity of the resulting estimates. This requires tools that support both accurate discharge detection and careful assessment of the identified MU spike trains.

In the version used for this tutorial (v0.2.0b1; https://www.giacomovalli.com/openhdemg/what%27s-new/#020-beta1), the *openhdemg* library provides an integrated set of tools for HDsEMG decomposition, MU inspection and cleaning through a graphical interface, and estimation of common synaptic input using different analytical approaches. However, given the precision required in MU cleaning for common synaptic input estimation, the workflow presented below uses the *openhdemg software* (v0.1.0b3; https://www.giacomovalli.com/openhdemg_software/), which is built on top of the *openhdemg* library and provides the most advanced graphical environment for detailed visual inspection, operator-guided refinement, and correction of MU discharge times.

### 6.1. Decomposition of HDsEMG signals into motor unit spike trains using *openhdemg*

HDsEMG signals were decomposed using the convolutive blind source separation approach implemented in *openhdemg*. This approach is based on the framework described by Negro et al. (2016a), in which multichannel EMG recordings are modelled as convolutive mixtures of MU spike trains, with the MU action potentials representing the finite-duration impulse responses through which each source is observed across the electrode array. To make the inverse problem more favorable for source separation, the recorded signals are temporally extended, so that the original convolutive mixture is reformulated as an instantaneous linear mixture of delayed observations (Holobar and Zazula, 2007b, Holobar and Zazula, 2007a). The extended observation matrix is then centered and whitened, reducing second-order correlations between channels and improving the conditioning of the source separation problem. MU sources are extracted iteratively using a fixed-point optimization procedure analogous to FastICA (Hyvärinen and Oja, 1997), where a nonlinear contrast function is used to identify sparse sources corresponding to candidate MU discharge trains. To reduce repeated convergence to the same source, each newly estimated separation vector is orthogonalized with respect to previously extracted separation vectors, reducing repeated convergence to the same source (Hyvärinen and Oja, 2000). For each extracted component, the corresponding separation vector is applied to the whitened extended EMG signals to obtain a continuous source estimate. This source estimate is then transformed by multiplying it by its absolute value, similar to the original formulation, but emphasizing large-amplitude peaks. Candidate discharge times are identified from the transformed signal by peak detection and then classified using two-class k-means clustering. The high-amplitude peak cluster is retained as the discharge train of the candidate MU, and its quality is quantified using the silhouette score. The separation vector is subsequently refined by recalculating it from the detected discharge times and repeating source estimation and peak classification while the silhouette score improves. Finally, only sources exceeding the selected silhouette threshold are retained as decomposed MUs.

The *openhdemg* implementation follows this general model while introducing practical and methodological adaptations. Whitening is performed using an eigenvalue-based dimensionality reduction step, in which low-variance components can be discarded according to a user-defined percentile threshold. The main decomposition parameters, including the extension factor, number of extraction iterations, contrast function, convergence tolerance, clustering initialization, maximum physiological discharge rate, minimum number of spikes, and silhouette threshold, are user-configurable. In addition, post-extraction separation vector refinement is implemented by recalculating the filter from the detected discharge times and iterating until the silhouette score improves, rather than using the coefficient of variation of the interspike intervals as the stopping criterion. Finally, the high-level decomposition pipeline includes optional band-pass filtering, power-line harmonic removal, exclusion of channels marked as poor quality, storage of all decomposition parameters, reconstruction of binary MU discharge trains, and post-processing removal of duplicate MUs. Thus, the implementation retains the physiological and mathematical basis of the Negro et al. (2016a) framework while providing a simple, reproducible, and configurable workflow for HDsEMG decomposition.

### 6.2. Visualization and cleaning of decomposed motor units using the *openhdemg software*

Following automatic decomposition, the identified MU spike trains must be visually inspected and, when necessary, manually refined before estimating common synaptic input (Avrillon et al., 2024, Del Vecchio et al., 2020, Afsharipour et al., 2020). This approach has been shown to be highly reliable across operators (Hug et al., 2021).

In the current workflow, for each MU, the operator inspected the estimated discharge times along with the corresponding source and discharge behavior, allowing potential errors in discharge-time detection to be identified. Typical corrections include removing spurious discharges, adding missed discharges when clearly detectable, excluding unreliable MUs, and checking for duplicated discharge patterns representing the same MU. MU cleaning was guided by both signal-based and physiological criteria. Signal-based criteria include the separation between discharge and noise clusters in the source estimate, quantified by the silhouette score (≥ 0.9), the consistency of the detected peaks, and the absence of duplicated discharge patterns. An additional important consideration for common synaptic input estimation was that the analysis should preferably be performed using MUs that discharge tonically throughout the selected analysis window. The inclusion of MUs with intermittent firing, recruitment or derecruitment events, or long silent periods can introduce slow fluctuations in the smoothed discharge rates that are not necessarily related to common synaptic input. These nonstationary changes may artificially increase or distort low-frequency oscillations, affecting the estimates. Physiological criteria include a plausible discharge rate range (Inglis et al., 2025), realistic interspike interval variability, and the absence of abrupt or non-physiological changes in firing behavior. It is important to note that this physiological assessment must always consider the task-imposed modulation, such as changes in force level or contraction dynamics, as well as any pathological condition of the participant that might affect the expected physiological discharge behavior.

Manual inspection and cleaning of MU spike trains is the most time-consuming step of the analysis. In the present workflow, the *openhdemg software* was used to streamline this process by combining rapid navigation and editing of discharge times using keyboard shortcuts, and simultaneous visualization of the source estimate and discharge behavior with real-time updates upon manual editing of the detected discharge times. Key quality controls, including the silhouette score, duplicate-unit detection, and MU action potential visualization, allowed the operator to quickly and reliably evaluate each MU using both quantitative indicators and physiological plausibility before retaining it for common synaptic input analyses.

### 6.3. Effect of motor unit cleaning on common synaptic input estimates

To demonstrate the effect of MU cleaning on common synaptic input estimates, we used the dataset containing 31 MUs described previously. Two versions of the same dataset were generated. In the first version, 8 MUs (13, 14, 16, 21, 22, 24, 29, and 30) were intentionally left insufficiently cleaned (but still with high silhouette values), while the remaining 23 MUs were fully cleaned (**Figure 11A**, left panel). In the second version, all 31 MUs were fully cleaned by an experienced operator (**Figure 11A**, right panel). Therefore, the two datasets differed only in the quality of the decomposition of these eight MUs. We then estimated common synaptic input using four of the approaches described in this tutorial: the common drive index (**Figure 11B**), principal component analysis of smoothed discharge rates (**Figure 11C**), pooled coherence (**Figure 11D**), and the proportion of common input index (**Figure 11E**). The two files used in the following examples, together with example codes to perform common synaptic input estimation using *openhdemg* as explained in this tutorial, are available at https://doi.org/10.6084/m9.figshare.32751243.

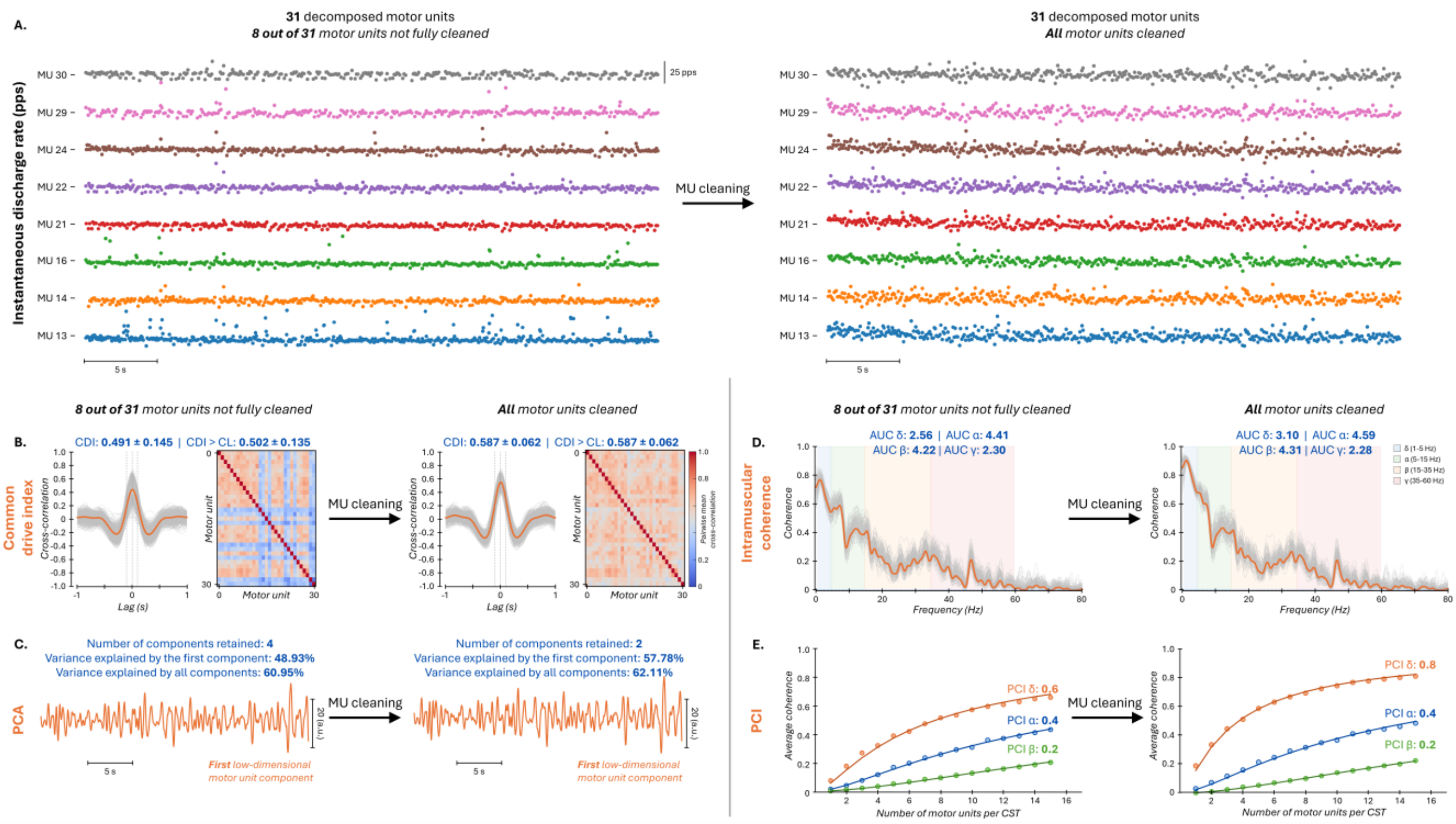

***Figure 11: Effect of motor unit cleaning on common synaptic input estimates.*** *(A) Example discharge rate of the motor unit used in the analysis. The left panel shows the dataset containing 8 insufficiently cleaned motor units (motor units 13, 14, 16, 21, 22, 24, 29, and 30), whereas the right panel shows the same dataset after all motor units were fully cleaned. The two datasets differed only in the quality of decomposition of these motor units. (B) Pairwise correlation matrices and common drive index estimates obtained from the insufficiently cleaned and fully cleaned datasets. (C) Principal component analysis of smoothed discharge rates for the two datasets. The insufficiently cleaned dataset resulted in a greater number of retained components and a lower percentage of variance explained*

*by the first component. (D) Pooled coherence estimates showing reduced coherence across frequencies in the insufficiently cleaned dataset, particularly within the delta band. (E) Proportion of common input (PCI) estimated from coherence-versus-CST-size curves. The delta-band PCI was substantially reduced in the insufficiently cleaned dataset, whereas alpha- and beta-band PCI estimates were comparatively less affected.*

The results revealed that all estimates were markedly affected by the presence of a relatively small number of insufficiently cleaned MUs, despite representing only 8 of the 31 identified MUs (approximately 25% of the MU population). For the common drive index, the thresholded common drive decreased from (0.59 ± 0.06) in the fully cleaned dataset to (0.50 ± 0.14) in the insufficiently cleaned dataset. Similarly, pooled coherence was systematically reduced, with the area under the coherence curve in the delta band decreasing from 3.10 to 2.56. The effect was even more pronounced for the proportion of common input index, where the delta-band PCI decreased from 0.8 to 0.6. In contrast, the PCI estimates obtained in the alpha and beta bands showed relatively small changes between conditions. Principal component analysis was also substantially affected by MU cleaning. In the insufficiently cleaned dataset, parallel analysis identified four components to be retained, with the first component explaining 49% of the variance of the smoothed discharge rates. After cleaning all MUs, the number of retained components decreased to two, while the variance explained by the first component increased to 58%. Therefore, decomposition inaccuracies may artificially increase the apparent dimensionality of the MU population while reducing the amount of variance explained by the dominant low-dimensional component. It is important to note that, despite the reduction in the amount of variance explained, the first principal component showed highly similar oscillatory behavior in both datasets. This finding indicates that the dominant low-dimensional component is reasonably robust to decomposition errors, although such errors can substantially alter estimates of dimensionality and explained variance.

Collectively, these results demonstrate that insufficiently cleaned MU spike trains can substantially alter estimates of common synaptic input obtained from both time-domain and frequency-domain approaches. The likely mechanism is that decomposition errors introduce spurious discharge events and missing discharges, effectively reducing the shared fluctuations that these methods aim to quantify. Importantly, the observed effects occurred even when the majority of MUs were properly cleaned, indicating that a relatively small number of inaccurate MU spike trains can bias population-level estimates. These findings highlight the importance of careful MU validation and manual refinement before estimating common synaptic input.

One of the main advantages of the current version of *openhdemg* is that HDsEMG decomposition, MU cleaning, and common synaptic input estimation can all be performed within the same integrated framework, either from the library or the software. This facilitates the implementation of a complete workflow, from the decomposition of HDsEMG signals to the extraction of physiologically meaningful estimates of common synaptic input, while maintaining full control over MU quality throughout the analysis pipeline.

## References

ADRIAN, E. D. & BRONK, D. W. 1928. The discharge of impulses in motor nerve fibres: Part I. Impulses in single fibres of the phrenic nerve. *J Physiol,* 66**,** 81-101.

AFSHARIPOUR, B., MANZUR, N., DUCHCHERER, J., FENRICH, K. F., THOMPSON, C. K., NEGRO, F., QUINLAN, K. A., BENNETT, D. J. & GORASSINI, M. A. 2020. Estimation of self-sustained activity produced by persistent inward currents using firing rate profiles of multiple motor units in humans. *J Neurophysiol,* 124**,** 63-85.

AHN, Y.-Y., BAGROW, J. P. & LEHMANN, S. 2010. Link communities reveal multiscale complexity in networks. *Nature,* 466**,** 761-764.

AMJAD, A. M., HALLIDAY, D. M., ROSENBERG, J. R. & CONWAY, B. A. 1997. An extended difference of coherence test for comparing and combining several independent coherence estimates: theory and application to the study of motor units and physiological tremor. *J Neurosci Methods,* 73**,** 69-79.

AVRILLON, S., HUG, F., BAKER, S. N., GIBBS, C. & FARINA, D. 2024. Tutorial on MUedit: An open-source software for identifying and analysing the discharge timing of motor units from electromyographic signals. *J Electromyogr Kinesiol,* 77**,** 102886.

BAKER, S. N. 2007. Oscillatory interactions between sensorimotor cortex and the periphery. *Current Opinion in Neurobiology,* 17**,** 649-655.

BAKER, S. N., OLIVIER, E. & LEMON, R. N. 1997. Coherent oscillations in monkey motor cortex and hand muscle EMG show task-dependent modulation. *J Physiol,* 501 ( Pt 1)**,** 225-41.

BAKER, S. N., PINCHES, E. M. & LEMON, R. N. 2003. Synchronization in monkey motor cortex during a precision grip task. II. effect of oscillatory activity on corticospinal output. *J Neurophysiol,* 89**,** 1941-53.

BALDISSERA, F., CAVALLARI, P. & CERRI, G. 1998. Motoneuronal pre-compensation for the low-pass filter characteristics of muscle. A quantitative appraisal in cat muscle units. *J Physiol,* 511 ( Pt 2)**,** 611-27.

BAWA, P. & STEIN, R. B. 1976. Frequency response of human soleus muscle. *J Neurophysiol,* 39**,** 788-93.

BIGLAND, B. & LIPPOLD, O. C. 1954. Motor unit activity in the voluntary contraction of human muscle. *J Physiol,* 125**,** 322-35.

BREMNER, F. D., BAKER, J. R. & STEPHENS, J. A. 1991. Correlation between the discharges of motor units recorded from the same and from different finger muscles in man. *J Physiol,* 432**,** 355-80.

CABRAL, H. V., COSENTINO, C., RIZZARDI, A., INGLIS, J. G., FUGLEVAND, A. J. & NEGRO, F. 2025a. Adaptations in common synaptic inputs to spinal motor neurons during grasping versus a less functional hand task. *Journal of Applied Physiology,* 139**,** 776-786.

CABRAL, H. V., CUDICIO, A., BONARDI, A., DEL VECCHIO, A., FALCIATI, L., ORIZIO, C., MARTINEZ-VALDES, E. & NEGRO, F. 2024a. Neural Filtering of Physiological Tremor Oscillations to Spinal Motor Neurons Mediates Short-Term Acquisition of a Skill Learning Task. *eneuro,* 11**,** ENEURO.0043-24.2024.

CABRAL, H. V., INGLIS, J. G., CUDICIO, A., COGLIATI, M., ORIZIO, C., YAVUZ, U. S. & NEGRO, F. 2024b. Muscle contractile properties directly influence shared synaptic inputs to spinal motor neurons. *The Journal of Physiology,* 602**,** 2855-2872.

CABRAL, H. V., INGLIS, J. G., POURREZA, E., DOS SANTOS, M. A., COSENTINO, C., O'REILLY, D., DELIS, I. & NEGRO, F. 2025b. A single low-dimensional neural component of spinal motor neuron activity explains force generation across repetitive isometric tasks. *iScience,* 28**,** 113483.

CASTRONOVO, A. M., NEGRO, F., CONFORTO, S. & FARINA, D. 2015. The proportion of common synaptic input to motor neurons increases with an increase in net excitatory input. *J Appl Physiol (1985),* 119**,** 1337-46.

CATTELL, R. B. 1966. The Scree Test For The Number Of Factors. *Multivariate Behav Res,* 1**,** 245-76.
CHEN, M. & ZHOU, P. 2016. A Novel Framework Based on FastICA for High Density Surface EMG Decomposition. *IEEE Trans Neural Syst Rehabil Eng,* 24**,** 117-27.
CHRISTAKOS, C. N., PAPADIMITRIOU, N. A. & ERIMAKI, S. 2006. Parallel neuronal mechanisms underlying physiological force tremor in steady muscle contractions of humans. *J Neurophysiol,* 95**,** 53-66.
CONWAY, B. A., HALLIDAY, D. M., FARMER, S. F., SHAHANI, U., MAAS, P., WEIR, A. I. & ROSENBERG, J. R. 1995. Synchronization between motor cortex and spinal motoneuronal pool during the performance of a maintained motor task in man. *J Physiol,* 489 ( Pt 3)**,** 917-24.
COSENTINO, C., CABRAL, H. V., DOS SANTOS, M. A., POURREZA, E., INGLIS, J. G. & NEGRO, F. 2026. Differential changes in the effective neural drive following new motor skill acquisition between vastus lateralis and medialis. *European Journal of Applied Physiology*.
COVER, T. M. 1999. *Elements of information theory*, John Wiley & Sons.
CUNNINGHAM, J. P. & YU, B. M. 2014. Dimensionality reduction for large-scale neural recordings. *Nature Neuroscience,* 17**,** 1500-1509.
DATTA, A. K. & STEPHENS, J. A. 1990. Synchronization of motor unit activity during voluntary contraction in man. *J Physiol,* 422**,** 397-419.
DE LA ROCHA, J., DOIRON, B., SHEA-BROWN, E., JOSIĆ, K. & REYES, A. 2007. Correlation between neural spike trains increases with firing rate. *Nature,* 448**,** 802-6.
DE LUCA, C. J. & ERIM, Z. 1994. Common drive of motor units in regulation of muscle force. *Trends Neurosci,* 17**,** 299-305.
DE LUCA, C. J. & FORREST, W. J. 1972. An electrode for recording single motor unit activity during strong muscle contractions. *IEEE Trans Biomed Eng,* 19**,** 367-72.
DE LUCA, C. J., LEFEVER, R. S., MCCUE, M. P. & XENAKIS, A. P. 1982. Control scheme governing concurrently active human motor units during voluntary contractions. *J Physiol,* 329**,** 129-42.
DEL VECCHIO, A., GERMER, C. M., ELIAS, L. A., FU, Q., FINE, J., SANTELLO, M. & FARINA, D. 2019. The human central nervous system transmits common synaptic inputs to distinct motor neuron pools during non-synergistic digit actions. *J Physiol,* 597**,** 5935-5948.
DEL VECCHIO, A., HOLOBAR, A., FALLA, D., FELICI, F., ENOKA, R. M. & FARINA, D. 2020. Tutorial: Analysis of motor unit discharge characteristics from high-density surface EMG signals. *J Electromyogr Kinesiol,* 53**,** 102426.
DEL VECCHIO, A., MARCONI GERMER, C., KINFE, T. M., NUCCIO, S., HUG, F., ESKOFIER, B., FARINA, D. & ENOKA, R. M. 2023. The Forces Generated by Agonist Muscles during Isometric Contractions Arise from Motor Unit Synergies. *J Neurosci,* 43**,** 2860-2873.
DERNONCOURT, F., AVRILLON, S., LOGTENS, T., CATTAGNI, T., FARINA, D. & HUG, F. 2025. Flexible control of motor units: is the multidimensionality of motor unit manifolds a sufficient condition? *The Journal of Physiology,* 603**,** 2349-2368.
DUCHATEAU, J. & ENOKA, R. M. 2011. Human motor unit recordings: origins and insight into the integrated motor system. *Brain Res,* 1409**,** 42-61.
ENOKA, R. M. & FARINA, D. 2021. Force Steadiness: From Motor Units to Voluntary Actions. *Physiology,* 36**,** 114-130.
FARINA, D. & NEGRO, F. 2015. Common synaptic input to motor neurons, motor unit synchronization, and force control. *Exerc Sport Sci Rev,* 43**,** 23-33.
FARINA, D., NEGRO, F. & DIDERIKSEN, J. L. 2014. The effective neural drive to muscles is the common synaptic input to motor neurons. *J Physiol,* 592**,** 3427-41.
FARINA, D., NEGRO, F., MUCELI, S. & ENOKA, R. M. 2016. Principles of Motor Unit Physiology Evolve With Advances in Technology. *Physiology (Bethesda),* 31**,** 83-94.

FARMER, S. F., BREMNER, F. D., HALLIDAY, D. M., ROSENBERG, J. R. & STEPHENS, J. A. 1993. The frequency content of common synaptic inputs to motoneurones studied during voluntary isometric contraction in man. *J Physiol,* 470**,** 127-55.
GALLET, C. & JULIEN, C. 2011. The significance threshold for coherence when using the Welch's periodogram method: Effect of overlapping segments. *Biomedical Signal Processing and Control,* 6**,** 405-409.
GALLOS, L. K., MAKSE, H. A. & SIGMAN, M. 2012. A small world of weak ties provides optimal global integration of self-similar modules in functional brain networks. *Proceedings of the National Academy of Sciences,* 109**,** 2825-2830.
GUTTMAN, L. 1954. Some necessary conditions for common-factor analysis. *Psychometrika,* 19**,** 149-161.
HAGBARTH, K. E. & YOUNG, R. R. 1979. Participation of the stretch reflex in human physiological tremor. *Brain,* 102**,** 509-26.
HAYTON, J. C., ALLEN, D. G. & SCARPELLO, V. 2004. Factor Retention Decisions in Exploratory Factor Analysis: a Tutorial on Parallel Analysis. *Organizational Research Methods,* 7**,** 191-205.
HECKMAN, C. J. & ENOKA, R. M. 2012. Motor unit. *Compr Physiol,* 2**,** 2629-82.
HOLOBAR, A. & ZAZULA, D. Gradient Convolution Kernel Compensation Applied to Surface Electromyograms. *In:* DAVIES, M. E., JAMES, C. J., ABDALLAH, S. A. & PLUMBLEY, M. D., eds. Independent Component Analysis and Signal Separation, 2007// 2007a Berlin, Heidelberg. Springer Berlin Heidelberg, 617-624.
HOLOBAR, A. & ZAZULA, D. 2007b. Multichannel Blind Source Separation Using Convolution Kernel Compensation. *IEEE Transactions on Signal Processing,* 55**,** 4487-4496.
HORN, J. L. 1965. A RATIONALE AND TEST FOR THE NUMBER OF FACTORS IN FACTOR ANALYSIS. *Psychometrika,* 30**,** 179-85.
HOTELLING, H. 1933. Analysis of a complex of statistical variables into principal components. *Journal of Educational Psychology,* 24**,** 417-441.
HOTELLING, H. 1936. Relations Between Two Sets of Variates. *Biometrika,* 28**,** 321-377.
HUG, F., AVRILLON, S., DEL VECCHIO, A., CASOLO, A., IBANEZ, J., NUCCIO, S., ROSSATO, J., HOLOBAR, A. & FARINA, D. 2021. Analysis of motor unit spike trains estimated from high-density surface electromyography is highly reliable across operators. *J Electromyogr Kinesiol,* 58**,** 102548.
HUG, F., AVRILLON, S., IBÁÑEZ, J. & FARINA, D. 2023a. Common synaptic input, synergies and size principle: Control of spinal motor neurons for movement generation. *The Journal of Physiology,* 601**,** 11-20.
HUG, F., AVRILLON, S., SARCHER, A., DEL VECCHIO, A. & FARINA, D. 2023b. Correlation networks of spinal motor neurons that innervate lower limb muscles during a multi-joint isometric task. *J Physiol,* 601**,** 3201-3219.
HYVÄRINEN, A. & OJA, E. 1997. A Fast Fixed-Point Algorithm for Independent Component Analysis. *Neural Computation,* 9**,** 1483-1492.
HYVÄRINEN, A. & OJA, E. 2000. Independent component analysis: algorithms and applications. *Neural Networks,* 13**,** 411-430.
INCE, R. A., GIORDANO, B. L., KAYSER, C., ROUSSELET, G. A., GROSS, J. & SCHYNS, P. G. 2017. A statistical framework for neuroimaging data analysis based on mutual information estimated via a gaussian copula. *Hum Brain Mapp,* 38**,** 1541-1573.
INGLIS, J. G., CABRAL, H. V., COSENTINO, C., BONARDI, A. & NEGRO, F. 2025. Motor unit discharge behavior in human muscles throughout force gradation: a systematic review and meta-analysis with meta-regression. *Journal of Applied Physiology,* 138**,** 1050-1065.
ISHIZUKA, N., MANNEN, H., HONGO, T. & SASAKI, S. 1979. Trajectory of group Ia afferent fibers stained with horseradish peroxidase in the lumbosacral spinal cord of the cat: three dimensional reconstructions from serial sections. *J Comp Neurol,* 186**,** 189-211.

JOLIFFE, I. & MORGAN, B. 1992. Principal component analysis and exploratory factor analysis. *Statistical Methods in Medical Research,* 1**,** 69-95.

KAISER, H. F. 1960. The Application of Electronic Computers to Factor Analysis. *Educational and Psychological Measurement,* 20**,** 141-151.

KIRKWOOD, P. A. 1979. On the use and interpretation of cross-correlations measurements in the mammalian central nervous system. *J Neurosci Methods,* 1**,** 107-32.

KIRKWOOD, P. A. & SEARS, T. A. 1978. The synaptic connexions to intercostal motoneurones as revealed by the average common excitation potential. *J Physiol,* 275**,** 103-34.

LAINE, C. M., MARTINEZ-VALDES, E., FALLA, D., MAYER, F. & FARINA, D. 2015. Motor Neuron Pools of Synergistic Thigh Muscles Share Most of Their Synaptic Input. *J Neurosci,* 35**,** 12207-16.

LAINE, C. M., NAGAMORI, A. & VALERO-CUEVAS, F. J. 2016. The Dynamics of Voluntary Force Production in Afferented Muscle Influence Involuntary Tremor. *Front Comput Neurosci,* 10**,** 86.

LAZAR, A. A. & PNEVMATIKAKIS, E. A. 2008. Faithful representation of stimuli with a population of integrate-and-fire neurons. *Neural Comput,* 20**,** 2715-44.

LECCE, E., CASOLO, A., NUCCIO, S., FELICI, F. & BAZZUCCHI, I. 2026. Analysis of motor units with high-density surface electromyography: methodological considerations and physiological significance. *Eur J Appl Physiol,* 126**,** 61-86.

LEMON, R. N. 2008. Descending pathways in motor control. *Annu Rev Neurosci,* 31**,** 195-218.

LEVINE, J., AVRILLON, S., FARINA, D., HUG, F. & PONS, J. L. 2023. Two motor neuron synergies, invariant across ankle joint angles, activate the triceps surae during plantarflexion. *The Journal of Physiology,* 601**,** 4337-4354.

LIDDELL, E. G. T. & SHERRINGTON, C. S. 1925. Recruitment and some other features of reflex inhibition. *Proceedings of the Royal Society of London. Series B, Containing Papers of a Biological Character,* 97**,** 488-518.

LIPPOLD, O. 1971. Physiological tremor. *Sci Am,* 224**,** 65-73.

MADARSHAHIAN, S., LETIZI, J. & LATASH, M. L. 2021. Synergic control of a single muscle: The example of flexor digitorum superficialis. *J Physiol,* 599**,** 1261-1279.

MAILLET, J., AVRILLON, S., NORDEZ, A., ROSSI, J. & HUG, F. 2022. Handedness is associated with less common input to spinal motor neurons innervating different hand muscles. *J Neurophysiol,* 128**,** 778-789.

MASUDA, T., MIYANO, H. & SADOYAMA, T. 1983. The propagation of motor unit action potential and the location of neuromuscular junction investigated by surface electrode arrays. *Electroencephalogr Clin Neurophysiol,* 55**,** 594-600.

MASUDA, T., MIYANO, H. & SADOYAMA, T. 1985. A surface electrode array for detecting action potential trains of single motor units. *Electroencephalogr Clin Neurophysiol,* 60**,** 435-43.

MCAULEY, J. H. & MARSDEN, C. D. 2000. Physiological and pathological tremors and rhythmic central motor control. *Brain,* 123 ( Pt 8)**,** 1545-67.

MCISAAC, T. L. & FUGLEVAND, A. J. 2008. Common synaptic input across motor nuclei supplying intrinsic muscles involved in the precision grip. *Exp Brain Res,* 188**,** 159-64.

MCMANUS, L., FLOOD, M. W. & LOWERY, M. M. 2019. Beta-band motor unit coherence and nonlinear surface EMG features of the first dorsal interosseous muscle vary with force. *J Neurophysiol,* 122**,** 1147-1162.

MERLETTI, R., FARINA, D. & GRANATA, A. 1999. Non-invasive assessment of motor unit properties with linear electrode arrays. *Electroencephalogr Clin Neurophysiol Suppl,* 50**,** 293-300.

MORI, S. 1975. Entrainment of motor-unit discharges as a neuronal mechanism of synchronization. *Journal of Neurophysiology,* 38**,** 859-870.

MUCELI, S., POPPENDIECK, W., HOLOBAR, A., GANDEVIA, S., LIEBETANZ, D. & FARINA, D. 2022. Blind identification of the spinal cord output in humans with high-density electrode arrays implanted in muscles. *Science Advances,* 8**,** eabo5040.

MUCELI, S., POPPENDIECK, W., NEGRO, F., YOSHIDA, K., HOFFMANN, K. P., BUTLER, J. E., GANDEVIA, S. C. & FARINA, D. 2015. Accurate and representative decoding of the neural drive to muscles in humans with multi-channel intramuscular thin-film electrodes. *J Physiol,* 593**,** 3789-804.

NEGRO, F. & FARINA, D. 2011a. Decorrelation of cortical inputs and motoneuron output. *J Neurophysiol,* 106**,** 2688-97.

NEGRO, F. & FARINA, D. 2011b. Linear transmission of cortical oscillations to the neural drive to muscles is mediated by common projections to populations of motoneurons in humans. *J Physiol,* 589**,** 629-37.

NEGRO, F. & FARINA, D. 2012. Factors influencing the estimates of correlation between motor unit activities in humans. *PLoS One,* 7**,** e44894.

NEGRO, F., HOLOBAR, A. & FARINA, D. 2009. Fluctuations in isometric muscle force can be described by one linear projection of low-frequency components of motor unit discharge rates. *J Physiol,* 587**,** 5925-38.

NEGRO, F., MUCELI, S., CASTRONOVO, A. M., HOLOBAR, A. & FARINA, D. 2016a. Multi-channel intramuscular and surface EMG decomposition by convolutive blind source separation. *J Neural Eng,* 13**,** 026027.

NEGRO, F., YAVUZ, U. & FARINA, D. 2016b. The human motor neuron pools receive a dominant slow-varying common synaptic input. *J Physiol,* 594**,** 5491-505.

NING, Y., ZHU, X., ZHU, S. & ZHANG, Y. 2015. Surface EMG decomposition based on K-means clustering and convolution kernel compensation. *IEEE J Biomed Health Inform,* 19**,** 471-7.

NORDSTROM, M. A., FUGLEVAND, A. J. & ENOKA, R. M. 1992. Estimating the strength of common input to human motoneurons from the cross-correlogram. *J Physiol,* 453**,** 547-74.

NUCCIO, S., GERMER, C. M., CASOLO, A., BORZUOLA, R., LABANCA, L., ROCCHI, J. E., MARIANI, P. P., FELICI, F., FARINA, D., FALLA, D., MACALUSO, A., SBRICCOLI, P. & DEL VECCHIO, A. 2024. Neuroplastic alterations in common synaptic inputs and synergistic motor unit clusters controlling the vastii muscles of individuals with ACL reconstruction. *J Appl Physiol (1985),* 137**,** 835-847.

O'REILLY, D. & DELIS, I. 2022. A network information theoretic framework to characterise muscle synergies in space and time. *J Neural Eng,* 19**,** 016031.

O'REILLY, D. & DELIS, I. 2024. Dissecting muscle synergies in the task space. *eLife,* 12**,** RP87651.

PEARSON, K. 1901. LIII. On lines and planes of closest fit to systems of points in space. *The London, Edinburgh, and Dublin Philosophical Magazine and Journal of Science,* 2**,** 559-572.

REUCHER, H., RAU, G. & SILNY, J. 1987. Spatial Filtering of Noninvasive Multielectrode EMG: Part I-Introduction to Measuring Technique and Applications. *IEEE Transactions on Biomedical Engineering,* BME-34**,** 98-105.

RODRIGUEZ-FALCES, J., NEGRO, F. & FARINA, D. 2017. Correlation between discharge timings of pairs of motor units reveals the presence but not the proportion of common synaptic input to motor neurons. *J Neurophysiol,* 117**,** 1749-1760.

ROSENBERG, J. R., AMJAD, A. M., BREEZE, P., BRILLINGER, D. R. & HALLIDAY, D. M. 1989. The Fourier approach to the identification of functional coupling between neuronal spike trains. *Prog Biophys Mol Biol,* 53**,** 1-31.

ROSENBERG, J. R., HALLIDAY, D. M., BREEZE, P. & CONWAY, B. A. 1998. Identification of patterns of neuronal connectivity--partial spectra, partial coherence, and neuronal interactions. *J Neurosci Methods,* 83**,** 57-72.

ROSSATO, J., AVRILLON, S., TUCKER, K., FARINA, D. & HUG, F. 2024. The volitional control of individual motor units is constrained within low-dimensional neural manifolds by common inputs. *The Journal of Neuroscience***,** e0702242024.

ROSSATO, J., TUCKER, K., AVRILLON, S., LACOURPAILLE, L., HOLOBAR, A. & HUG, F. 2022. Less common synaptic input between muscles from the same group allows for more flexible coordination strategies during a fatiguing task. *J Neurophysiol,* 127**,** 421-433.
SEARS, T. A. & STAGG, D. 1976. Short-term synchronization of intercostal motoneurone activity. *J Physiol,* 263**,** 357-81.
SHERRINGTON, C. S. 1925. Remarks on some aspects of reflex inhibition. *Proceedings of the Royal Society of London. Series B, Containing Papers of a Biological Character,* 97**,** 519-545.
THOMPSON, C. K., NEGRO, F., JOHNSON, M. D., HOLMES, M. R., MCPHERSON, L. M., POWERS, R. K., FARINA, D. & HECKMAN, C. J. 2018. Robust and accurate decoding of motoneuron behaviour and prediction of the resulting force output. *J Physiol,* 596**,** 2643-2659.
VALLI, G., RITSCHE, P., CASOLO, A., NEGRO, F. & DE VITO, G. 2024. Tutorial: Analysis of central and peripheral motor unit properties from decomposed High-Density surface EMG signals with openhdemg. *Journal of Electromyography and Kinesiology,* 74**,** 102850.
VELICER, W. F. 1976. Determining the number of components from the matrix of partial correlations. *Psychometrika,* 41**,** 321-327.
VELICER, W. F., EATON, C. A. & FAVA, J. L. 2000. Construct Explication through Factor or Component Analysis: A Review and Evaluation of Alternative Procedures for Determining the Number of Factors or Components. *In:* GOFFIN, R. D. & HELMES, E. (eds.) *Problems and Solutions in Human Assessment: Honoring Douglas N. Jackson at Seventy.* Boston, MA: Springer US.
WELCH, P. 1967. The use of fast Fourier transform for the estimation of power spectra: A method based on time averaging over short, modified periodograms. *IEEE Transactions on Audio and Electroacoustics,* 15**,** 70-73.
ZWICK, W. R. & VELICER, W. F. 1986. Comparison of five rules for determining the number of components to retain. *Psychological Bulletin,* 99**,** 432-442.